\documentclass[aps,pre,reprint,showpacs,amssymb,showpacs,twocolumns,amsmath,superscriptaddress]{revtex4-1}

\usepackage[dvips]{epsfig,graphicx}
\usepackage{amsmath,amssymb,amsfonts,wasysym,color}

\def\be{\begin{equation}}
\def\ee{\end{equation}}
\def\la{\langle}
\def\ra{\rangle}

\def\Cas{\mathrm{C}}
\def\bulk{\mathrm{bulk}}

\def\ex{\mathrm{ex}}
\def\kB{k_{\mathrm{B}}}

\def\C{{\mathcal C}}

\def\rme{{\rm e}}
\def\rmd{{\rm d}}
\newcommand\reff[1]{(\ref{#1})}
\def\bea{\begin{eqnarray}}
\def\eea{\end{eqnarray}}

\def\sfh{\hat\vartheta}

\begin{document}
\title{
Critical Casimir forces for  Ising films with variable
 boundary fields
}
\author{O.~Vasilyev}
\affiliation{Max-Planck-Institut f{\"u}r Intelligente Systeme, Heisenbergstr.~3, D-70569 Stuttgart, Germany}
\affiliation{Institut f{\"u}r Theoretische und Angewandte Physik, Universit{\"a}t Stuttgart, Pfaffenwaldring 57, D-70569 Stuttgart, Germany}

\author{A.~Macio\l ek}
\affiliation{Max-Planck-Institut f{\"u}r Intelligente Systeme, Heisenbergstr.~3, D-70569 Stuttgart, Germany}
\affiliation{Institut f{\"u}r Theoretische und Angewandte Physik, Universit{\"a}t Stuttgart, Pfaffenwaldring 57, D-70569 Stuttgart, Germany}
\affiliation{Institute of Physical Chemistry,  Polish Academy of Sciences, Kasprzaka 44/52, PL-01-224 Warsaw, Poland}

\author{S.~Dietrich}
\affiliation{Max-Planck-Institut f{\"u}r Intelligente Systeme, Heisenbergstr.~3, D-70569 Stuttgart, Germany}
\affiliation{Institut f{\"u}r Theoretische und Angewandte Physik, Universit{\"a}t Stuttgart, Pfaffenwaldring 57, D-70569 Stuttgart, Germany}
\date{\today}

\begin{abstract}
Monte Carlo simulations based   on an integration scheme for free energy
differences is used to compute critical Casimir
forces for three-dimensional  Ising 
films with various  boundary fields. We study the scaling behavior of the critical
Casimir force, including  the scaling variable related to the boundary fields.
Finite size corrections to scaling are taken into account.
We pay  special attention to that range of surface field strengths within
 which the force changes from  repulsive to attractive
 upon increasing  the temperature.
Our data   are compared with other results available in the literature.

\end{abstract}
\pacs{05.50.+q, 05.70.Jk, 05.10.Ln, 68.15.+e}


\maketitle

\section{Introduction}
\label{sec:intr}
Forces induced by thermal fluctuations  can be very sensitive to 
tiny changes in  temperature.
This is exemplified by  effective forces  arising  between two surfaces 
confining a fluid
 close to its critical point, for which
a slight variation in temperature can lead to  pronounced changes 
in their  range and  magnitude.
The universal features of these so-called critical Casimir forces are captured  by  scaling
functions \cite{FdG,krech:99:0,dantchev}; they have been 
 studied theoretically and experimentally
for systems belonging to  the bulk universality classes of 
 the $XY$ and the  Ising   model \cite{krech:99:0,dantchev,gambassi}.
The $XY$ model describes quantum fluids, such as  liquid $^4$He close to its normal-superfluid 
phase transition
or a  $^3$He-$^4$He mixture close to its tricritical point, whereas, e.g., a  classical binary liquid  mixture near its  
demixing point or a simple fluid close to a liquid-gas  critical point belong to the Ising universality class.

In the systems studied so far  experimentally,
the measured critical Casimir forces
have been  either attractive or repulsive throughout  the 
{\it whole} temperature range. (The addition of salt to a critical oil-water mixture
presents a notable exception in that, under favorable conditions, on route to the critical demixing point
the sign of the critical Casimir force can change twice \cite{soft_matter} 
due to a coupling between the noncritical charge density and the critical order parameter field.)
 Here, we  investigate simple  systems
 which provide  the possibility 
of changing the sign of the critical Casimir forces upon varying the temperature.
Analytic studies and computer simulations indicate
that the sign of the critical Casimir
force 
is determined by the properties of the confining surfaces, i.e., 
 by the boundary conditions (BCs) which they impose
on the fluctuations of the order parameter characterizing the underlying second-order 
phase transition.
Indirect measurements of the Casimir scaling function, inferred from wetting films of superfluids \cite{garcia,garcia1} and
of classical binary liquid mixtures  \cite{pershan,rafai}, 
are consistent with these predictions.
For  pure $^4$He  one has symmetric
 Dirichlet-Dirichlet $(O,O)$ BCs because 
the quantum mechanical wave function of the superfluid
state
vanishes at both confining interfaces. This  gives 
 rise to  attractive critical Casimir forces
\cite{EPL,PRE,hucht,kardar,LGW-MF}. For wetting films of $^3$He-$^4$He mixtures,  $(+,O)$ BCs 
are realized because due to quantum mechanical effects a $^4$He-rich layer forms near the solid-liquid interface
and favors the superfluid phase giving rise to the so-called
 surface transition \cite{LGW-MF,MD_EPL}; $(+)$ indicates a symmetry-breaking BC with
 the surface completely ordered.
Upon reaching  the surface transition the  superfluid order parameter becomes
  nonzero at the solid surface whereas it 
vanishes at the fluid-vapor interface of the wetting film. These
asymmetric BCs give rise to a repulsive Casimir force.
Measurements for  wetting films of certain classical binary liquids  mixtures 
have been found to be in
agreement with  $(+ -)$ BCs corresponding to a strong opposing preferential adsorption 
of the two species of the mixture 
at the two confining surfaces \cite{krech,EPL,PRE,upton}. 
Within the framework of an  Ising magnet  (which is equivalent to the lattice model of a binary mixture) or within the 
continuum field theory for the order parameter, this amounts
 to the presence of strong antagonistic symmetry-breaking surface fields
${\bar H}_1$ and ${\bar H}_2$  which  couple linearly to the order parameter 
 and give rise to  repulsive critical Casimir forces.
Direct evidences for critical Casimir forces have been provided by  studying the Brownian motion of  
a  single colloidal  particle near   a flat substrate surface and 
immersed  in the binary liquid mixture of  water and lutidine
 \cite{Hertlein,PRE_nature}. The experimental results 
for the cases in which the colloid
and the substrate surface preferentially adsorb the same species of the mixture
 are consistent with  $(++)$ or 
$(--)$ BCs, whereas  for cases  in which 
the particle and the surface
 preferentially adsorb different species of the mixture the results
agree with the occurrence of $(-+)$ or $(+-)$ BCs. Whereas the theoretical and experimental
 understanding of critical Casimir forces in the presence of strong or
vanishing   surface fields 
has reached a mature level, here we set out to study the influence of variable weak surface fields.

 Dirichlet and  $(\pm)$  BCs  are the
renormalization-group  fixed-point boundary conditions corresponding
 to the so-called ordinary surface universality class $(O)$ and the normal transition 
surface universality class, respectively 
\cite{binder:83:0,diehl:86:0,diehl:97}. 
The ordinary transition corresponds to the bulk  phase transition  
occurring in the absence of  surface fields and with   a reduced tendency to order 
at the surface. Within  a mean field picture the latter
 is described by a  surface scaling field $c$ so that $1/c$  plays the role of  an extrapolation 
length of the order parameter profile; $c=\infty$ defines the ordinary transition
fixed point. The normal transition occurs for
systems with strong surface fields and which exhibit  a reduced tendency to order if these surface
fields are switched off. The normal transition  is defined by 
the fixed point (${\bar H}_1=\infty,c=\infty$).
As indicated by the nomenclature the normal transition is the generic  situation for a fluid.
In a spin model as discussed below  $H_1={\bar H}_1/J$ is dimensionless with $J$ as an interaction constant (see below).

Near the ordinary transition there is a single linear scaling field $g_{1}=H_1/c^y$
associated  with the surface 
field of strength $H_1$ and the surface enhancement parameter $c$  \cite{diehl:97}.
The scaling exponent is $y=\left(\Delta_1^{sp}-\Delta_1^{ord}\right)/\Phi$, 
 where  $\Delta^{sp,ord}_{1}$ are the surface counterparts of the bulk gap exponent $\Delta$
 and $\Phi$ is a  crossover exponent \cite{binder:83:0,diehl:86:0,diehl:97}. 
For  the  three-dimensional $(D=3)$  Ising model one has  $\Delta^{ord}_{1}\simeq 0.46(2)$  \cite{GZ},
 $\Delta_1^{sp}\simeq 1.05$ \cite{diehl:86:0}, $\Phi\simeq 0.68$ \cite{diehl:86:0}, and $y\simeq 0.87$; $\nu \simeq 0.63$ \cite{PV}
is the critical exponent of the bulk correlation length $\xi_b=\xi_0^{\pm}|t|^{-\nu}$ with the reduced temperature
$t=
(\beta_{c}-\beta)/\beta = (T-T_{c})/T_{c}$;  $\pm $ corresponds to $t\gtrless 0$.
 The corresponding  surface  scaling variable   can be chosen as 
$(\xi_0^+)^{-y}g_{1}|t|^{-\Delta^{ord}_{1}}$.
For $t\to 0$ the scaling variables tend to their fixed point values,
 and the scaling functions
assume asymptotic forms corresponding to 
the respective fixed points  \cite{diehl:86:0,diehl:97}.
The scaling variable $|(\xi_0^+)^{-y}g_{1}|t|^{-\Delta^{ord}_{1}}|^{\nu/\Delta^{ord}_1}$ is proportional to the ratio
$\xi_b/\ell_1$  between the true bulk correlation length $\xi_b$ 
 and the length $\ell_1$ introduced by the scaling field $g_1$:
\begin{equation}
\label{eq:length}
\ell_1 = \xi_0^+|(\xi_0^+)^{-y}g_1|^{-\nu/\Delta_1^{ord}}.
\end{equation} 
The fixed point dominated  critical regions correspond to 
either the divergence 
( $(+)$ or $(-)$ fixed-point BCs) or
the vanishing  ( $(O)$  fixed-point BCs)  of this ratio.
The length $\ell_1$ corresponds 
to the range of distances from the
surface  within  which the order parameter profile responds linearly to the presence of a surface  field $H_1$
\cite{ritschel_czerner,ciach_maciolek_stecki}.
(A precise definition of $\ell_1$ will be provided below.)

Depending  on the interplay between  $\ell_1$ and  
 the length
 $\ell_c = (\xi_0^+c)^{-\nu/\Phi}$ 
 associated with the surface enhancement parameter $c$
one finds various asymptotic regimes for  the  short-distance behavior $z \ll \xi_b$ of 
the order parameter profile 
\cite{diehl:97,ritschel_czerner,ciach_ritschel}.
At the bulk  critical point one has 
$\phi_{cri}(z) \sim g_1z^{\kappa}$  for distances  $\ell_c \ll z \ll \ell_1$ from the surface 
with $\kappa=(\Delta_1^{ord}-\beta)/\nu$ and 
 $\phi_{cri}(z) \sim z^{-\beta/\nu}$ 
for distances $\ell_c \ll \ell_1\ll z$ 
from the surface.
We note that near the ordinary transition fixed-point (i.e., large $c$)  the length $\ell_c$ is small
whereas the length $\ell_1$ can be large or small.
 Within mean field theory one has 
$\kappa=0$ due to $\Delta_1^{ord}(D=4)=1/2$ and $\nu(D=4)=1/2$ whereas
 one has  $\kappa(D=3)\simeq 0.23$ \cite{ritschel_czerner}
and $(\beta/\nu) (D=3) \simeq 0.52$ \cite{PV}. 
Consequently, the critical order parameter (OP) profile 
turns out to be a nonmonotonic function of $z$.
For  $z\ll \ell_1$  the OP increases upon increasing $z$, at $z\simeq \ell_1$ it reaches 
a maximum and only for 
$z\gg \ell_1$  the universal ''normal`` fixed-point behavior, i.e., the decay of the
OP occurs \cite{ritschel_czerner}. 
Accordingly the position of this maximum can serve as a definition for the length
 $\ell_1$ \cite{ritschel_czerner,ciach_maciolek_stecki}.
With increasing surface field strength the surface-near regime with the 
aforementioned increase 
$\sim z^{\kappa}$ of the OP  becomes narrower, 
and eventually for $H_1\to \infty$ the length scale $\ell_1$ goes to zero, such that this regime
disappears  and the normal transition  behavior  $\sim z^{\beta/\nu}$ is attained throughout.

For the $2D$ Ising model on the square lattice with lattice constant $a$,
the length $\ell_1$ has been extracted  from 
 an exact result for  the scaling function of the OP profile below and above $T_c$; the profile at $T_c$
has not been reported. 
In the case that  the exchange  coupling  between  spins in the surface row is the same as in the bulk,
the OP scaling function   depends  on the scaling variable 
$\xi_b/{\hat l}_1$ with ${\hat l}_1=(a/2)\tanh(K)/(\tanh {\bar h_1})^2$, where $K=J/(k_BT)$ is the dimensionless reduced exchange coupling
between Ising spins  and ${\bar h_1}={\bar H_1}/(k_BT)$ \cite{bariev}.
Thus  for weak surface fields and in the limit 
$K\to K_c$,  ${\hat l}_1(K_c)= (a/2)K_c^{-2}\tanh(K_c)/H_1^2 = 1.066(4) a {H_1}^{-2}= 1.879(0)\xi_0^+H_1^{-2}$, 
where $K_c=0.5\ln (1+\sqrt 2)\simeq 0.44$ is the critical coupling and $\xi_0^+=a/(4K_c)$. 
This is in line with Eq.~(\ref{eq:length}) due to $\nu(D=2)=1$ and $\Delta_1^{ord}(D=2)=1/2$.
 Examination of the OP profiles for $T\to T_c$ shows that the maximum
occurs at $z_{max}\simeq  1.5 {\hat l}_1$ which implies  $\ell_1\simeq 2.8 \xi_0^+H_1^{-2}$.

Studies  of systems belonging 
to the Ising universality class  \cite{MCD,MEW,MDB,MMD,abraham_maciolek} showed that near bulk criticality
the presence of the  length scale $\ell_1$ 
has important consequences for finite-sized systems such as slabs of thickness $L$.
For these systems the relevant
lengths are  the bulk correlation length $\xi_b$,  the distance $L$ 
between the two confining surfaces which exert fields $H_1$ and $H_2$, and 
the corresponding lengths $\ell_1$ and $\ell_2$.
The asymptotic critical region, associated with  $(+)$ or $(-)$ fixed-point
boundary conditions at the surfaces $i=1, 2$,  corresponds to 
$L \gg \ell_{i}$,  whereas corrections
proportional to $\ell_{i}/L$ 
are expected to be relevant for $L\simeq \ell_{i}$.
In the crossover regime the critical properties of the confined systems are particularly 
sensitive to the values of the surface fields, i.e., 
whether  one or both length scales 
$\ell_i$ become comparable to or even larger than the distance $L$,  together with 
$L, \ell_i \ll \xi_b$.
For example, in 
films with identical surface fields, i.e., $H_1=H_2$ and $ \ell_1=\ell_2$, 
at bulk criticality and for weak surface fields
the  order parameter  profile exhibits two symmetric maxima at 
$z\simeq \ell_1$ and $z\simeq L-\ell_1$; for even weaker 
fields so that  $\ell_1\simeq L$ these maxima merge into a single one at midpoint $z=L/2$
\cite{MCD,MEW}. Concomitantly the critical Casimir amplitude as a function of the
surface field $H_1$, i.e.,
the critical Casimir force 
at the bulk critical temperature, exhibits  a maximum absolute value at $L \simeq \ell_1$  \cite{MCD} . 

For symmetric surfaces, the effect of variation of  the amplitude 
of $H_1$  on the temperature dependence of the critical Casimir force,
 i.e., the crossover behavior between the ordinary and normal surface
 universality classes,
was studied within the two-dimensional $(2D)$ Ising model by using the 
quasi-exact numerical density-matrix renormalization-group method \cite{MDB}
 and within continuum  mean-field theory \cite{MMD}.
For  $L/\ell_1 \sim 1$  these results  show  strong deviations of the force scaling
function  from its universal fixed-point  behavior 
such as the  occurrence
of two minima, one above and one below   $T_c$, but  {\it no}
 change in sign  as the
 temperature is varied.
It turns out that only  strongly asymmetric  surface fields can lead to, even
 multiple, 
sign changes of the critical Casimir forces 
upon varying the temperature. This has been  demonstrated rigorously
for  $2D$ Ising films \cite{abraham_maciolek}, within mean-field
theory  for  the same geometry  \cite{MMD},
and it was supported by our  preliminary results from
Monte Carlo simulations of  simple cubic Ising slabs \cite{MMD}.
Further evidence has been  provided by  Monte Carlo simulations of the improved
{\it Blume-Capel} model  in the film geometry \cite{hasen2}.
The Blume-Capel model has a  second-order phase transition which also belongs to the $3D$
Ising  universality 
class.
It offers the opportunity that a careful choice of the interaction  parameters of this model
allows one  to eliminate  leading corrections to 
finite-size scaling  (see also Ref.~\cite{parisen}). As it will be discussed below, 
controlling finite-size corrections is essential for inferring the 
scaling functions of critical Casimir forces from 
Monte Carlo simulation  data. 
In the following, we shall present a   Monte Carlo simulation study of 
the critical Casimir forces for the
$3D$ {\it Ising} model  in a slab geometry with freely variable surface fields applied at its
 bottom and top surfaces.  Our scan of the parameter space extends the one presented in Ref.~\cite{hasen2}. As mentioned above, in Ref.~\cite{MMD} certain preliminary results of
this study
were reported  together with a detailed continuum   mean-field analysis.

The analytic results and the simulation data
 for  the scaling functions of the critical
Casimir forces for  weak surface fields can be probed
experimentally and they offer  application perspectives 
for soft matter systems such as  tuning
the properties of colloidal suspensions.
A first attempt to investigate  experimentally the effects  of  gradual
changes in the  properties   of confining surfaces on critical Casimir forces was made 
 recently by studying   colloids  suspended in a critical
 mixture of water and lutidine \cite{NHB}.
These experiments have    demonstrated
 the ability   to continuously
tune the order parameter boundary conditions at  the confining
surfaces. This was achieved by a chemical treatment of a solid substrate
such that it produces a spatial gradient of the adsorption preference for lutidine
and water molecules. Depending on the position of a single disolved colloidal particle 
 at this structured  surface a smooth transition from attractive
 to repulsive critical Casimir forces was found. 
However, these experimental observations have not yet  been cast 
into a universal scaling function of the critical Casimir potential which has to change sign as 
function of the effective surface field.

Our presentation is organized as follows. In Sec.~\ref{sec:theory} we
introduce our model, define the range of parameters for which we perform our computations,
and briefly present the relevant theoretical background. In Sec.~\ref{sec:method}
we describe the numerical method employed in order to infer  the scaling functions of the critical Casimir forces
from the MC simulation data. In Sec.~\ref{sec:corr} we discuss corrections to scaling 
which we take  into account in order to obtain 
data collapse signalling scaling.
Section \ref{sec:res} contains our results. We provide a summary and conclusions in Sec.~\ref{sec:concl}.

%
 

\section{Model and  Theoretical Background}
\label{sec:theory}
In the spirit of the universality of critical phenomena we study the simplest
representative of the $3D$ Ising universality class, i.e.,  the three-dimensional  Ising model
  defined 
on  a simple cubic lattice. We consider 
  a slab geometry. The dimensionless volume  of the system is
 $L_{x} \times L_{y} \times L_{z}$
where $L_{x}=L_{y} \gg L_{z}$ and $A=L_x \times L_y$
 with periodic BCs along  the  $x$ and $y$ directions.  
Each lattice site $(x,y,z) $ with
$1 \le x \le L_{x}, 1 \le y \le L_{y}, 1 \le z \le L_{z}$  and lattice constant 1 is occupied by
a spin $s_{x,y,z}=\pm 1$.  
The Hamiltonian of the Ising model with surface fields is
\be
\label{eq:Ham}
\frac{{\cal H}}{J} = -\sum_{\la {\rm nn} \ra}  s_{x,y,z}  s_{x',y',z'}
+H_{1}^{-} \sum_{ x,y }s_{x,y,1}+
H_{1}^{+}\sum_{ x,y }s_{x,y,L_{z}}, 
\ee
where $J>0$ is the spin-spin  interaction constant,
$\bar H_{1}^{-}=H_{1}^{-}J$ and $\bar H_{1}^{+}=H_{1}^{+}J$ are the values of the surface boundary fields
acting on the spins in the   bottom and in the top layer, respectively.
The sum $\la {\rm nn} \ra$ is taken  over all nearest-neighbor pairs
of sites on the lattice and the sum $x,y$ corresponding to the  boundary fields
is taken over the  top and the bottom layer. Here we do not consider a bulk field. In the following
temperatures, the surface fields, and energies are measured in units of $J$;
the inverse critical temperature is 
$\beta_c=0.2216544(3)$~\cite{RZW}.

%
\begin{figure}[h]
\includegraphics[width=0.5\textwidth]{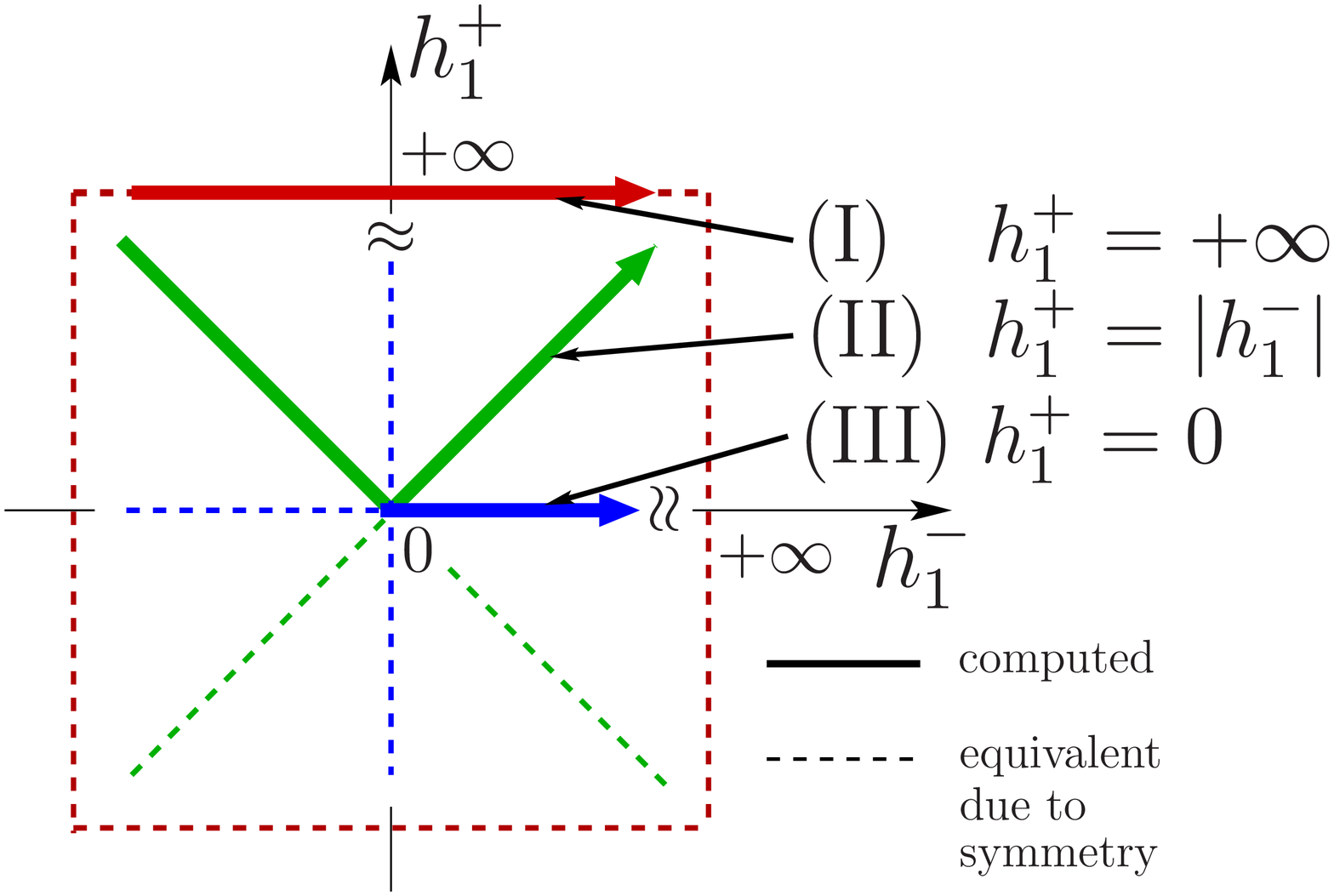}
\caption{
The parameter space spanned by the  scaling variables 
$ h_{1}^{+}$ and $ h_{1}^{-}$  
corresponding to the top and bottom surface fields $H_1^+$ and $H_1^-$, respectively (see Eq.~(\ref{eq:sch})).
We investigate the  following paths: (I) $h_{1}^{+}=\infty$ (red line)
 corresponding to  an infinitely strong surface field  $H_1^+$,
(II) $h_{1}^{+}=|h_{1}^{-}|$ (green line), (III) $h_{1}^{+}=0$ (blue line).
Dashed lines of corresponding colors denote trajectories, which
are equivalent due to the exchange symmetry $h_{1}^{+} \leftrightarrow h_{1}^{-}$.
Since Eq.~(\ref{eq:Ham}) does not contain a bulk field there is in addition the symmetry
$(h_{1}^{+},h_{1}^{-})\leftrightarrow (-h_{1}^{+},-h_{1}^{-})$.
}
\label{fig:fig1}
\end{figure}
%

For a fixed width  $L_z$ and a fixed aspect ratio $\rho=L_z/L_x=L_z/L_y$ of the slab,  the thermodynamic  
state of the system is
characterized  by three parameters: $t,H_{1}^{-}$, and $H_{1}^{+}$.
Based on finite-size scaling arguments, for the present system 
Fisher and Nakanishi \cite{fisher_nakanishi} proposed  
the following convenient  
 scaling variables associated with the  surface fields:
\begin{eqnarray}
\label{eq:sch}
 h_{1}^{\pm} := H_{1}^{\pm}L_z^{\Delta^{ord}_{1}/\nu}&=&(c\xi_0^+)^y
\left[ L_z/(\ell_1^{\pm}/\xi_0^+)\right]^{\Delta^{ord}_{1}/\nu} \\ \nonumber 
&\stackrel{D=3}{=}&(c\xi_0^+)^{0.87}
\left[ L_z/(\ell_1^{\pm}/\xi_0^+)\right]^{0.73};
\end{eqnarray} 
$\ell^-_1$ and $\ell^+_1$ correspond to bottom and the top surface, respectively.

Here we study  the 
following three trajectories (see Fig.~\ref{fig:fig1}):

\begin{itemize}
\item[(I)] $ h_{1}^{+}=\infty$,  \hspace{0.11cm} an infinitely strong top surface field. 

\item[(II)] $ h_{1}^{+}=|h_{1}^{-}|$,  finite  symmetric  and antisymmetric surface fields. 

\item[(III)] $ h_{1}^{+}=0$,   \hspace{0.4cm} free  boundary conditions at the top surface.

\end{itemize}
In the simulations,  case (I) is realized by
fixing all spins in the top layer $z=L_z$ at the value $+ 1$.
For  finite surface fields 
it is convenient to replace the surface field applied at the 
top (bottom)  surface  of the slab by having this surface layer being linked via  modified bonds 
to  spins located in an extra layer  $z=0$ ( $z=L_z+1$)
with the interaction $-H_{1}^{-} \sum_{ x,y }s_{x,y,0}s_{x,y,1}$
($-H_{1}^{+} \sum_{ x,y }s_{x,y,L_z}s_{x,y,L_z+1}$);  the spins in the extra layer
 $z=0$ ($z=L_z+1$) are fixed at the same value $+1$ 
for all $x, y$.
In practice, a  surface field,
which  is  finite but strong enough 
to lead to  a saturation of the data, can be  used to mimic the action 
of an infinite surface field.
For instance,  for    $|h_{1}^{+}|>100$ we do not observe any variation of our data as function of $h_1^+$.

By construction, for  all three cases there is only  one scaling variable associated with  the two
surface fields. In the following  we use the notation  $H_{1}\equiv H_{1}^{-}$
and $h_{1} \equiv h_{1}^{-}$. This fixes  the top surface field 
$H_{1}^{+}$  in accordance with (I) - (III). 
The plane 
of parameters $(h_{1}^{-}=h_{1},h_{1}^{+})$ is shown in Fig.~\ref{fig:fig1}. 
We note that
 the 
 cases (I) and  (II) with symmetric fields  coincide at the point $(\infty, \infty)$, 
the  cases (II) with symmetric fields and (III)  coincide for $(0,0)$, 
and finally for the  cases (I)  and (III) the  point  $(0, \infty)$ coincides with  $(\infty,0)$.

We have computed the critical  Casimir forces for a selection of
parameters from sets  
corresponding to the cases (I), (II), and (III)  which in Fig.~\ref{fig:fig1} are denoted by  solid lines.
Points in the plane $(h_{1}^{-},h_{1}^{+})$ corresponding to the cases
 (I), (II), and  (III)  which 
are equivalent
due to the exchange symmetry
$h_{1}^{-} \leftrightarrow h_{1}^{+}$  are indicated
by dashed lines. Due to the absence of  a bulk field there is also the symmetry
$(h_{1}^{+},h_{1}^{-})\leftrightarrow (-h_{1}^{+},-h_{1}^{-})$.

For large areas  $A$, the total free energy
$F(\beta,H_{1}
^+,H_1^-,L_z,A)$ of the  film  of thickness 
$L_z$ can be written as 
\begin{eqnarray}
\label{free_en1}
&& \frac{F(\beta,H_{1}^+,H_1^-,L_z,A)}{A} \equiv L_zf(\beta, H_{1}^+,H_1^-,L_z) \\ \nonumber
& = & L_z f^\bulk(\beta) + \beta^{-1} f^\ex(\beta,H_{1}^+,H_1^-,L_z),
\end{eqnarray}
 where $f^\bulk(\beta)$ is the bulk free energy density at a given temperature.
The excess free energy $f^\ex$ per  area  contains two $L_z$-independent surface  contributions
in addition to the finite-size contribution
$f^\ex(\beta,H_{1}^+,H_1^-,L_z)-f^\ex(\beta,H_{1}^+,H_1^-,\infty)$ which vanishes for $L_z\to \infty$.
The 
$L_z$-dependence of the latter
gives rise to the critical  Casimir force  $f_\Cas$ per unit area $A$ 
and in units of $\kB T \equiv \beta^{-1}$:
\be
\label{eq:def}
f_\Cas(\beta,H_{1}^+,H_{1}^-,L_z)\equiv - \partial f^\ex(\beta,H_{1}^+,H_{1}^-,L_z)/\partial L_z,
\ee
with the  bottom surface field 
$H_{1}^{-}=H_{1}$ and  the upper surface field $H_{1}^{+}=\{\infty,|H_{1}^{-}|,0 \}$, 
in accordance with (I), (II), and (III), respectively.  

For a lattice (lattice quantities are denoted by a  ``hat''
$\,\hat{}\,$), the derivative in Eq.~(\ref{eq:def})
is replaced by a finite difference and $\hat f_\Cas(\beta,L)$ is given by 
\be 
\label{eq:force}
\hat f_\Cas(\beta,H_{1},L,A)
\equiv - \frac{\beta \Delta \hat F(\beta,H_{1},L,A)}{A}
+ \beta \hat f^\bulk(\beta)\,,
\ee
with  the free energy difference 
$\Delta \hat F(\beta,H_{1},L,A)= \hat F(\beta,H_{1},L+\frac{1}{2},A)-
\hat F(\beta,H_{1},L-\frac{1}{2},A)$.
In these three expressions the thickness  $L=L_{z}-\frac{1}{2}$ is half-integer, so that the rhs
 is expressed
via the free energy difference for slabs of 
integer thicknesses  $L_{z}=L+\frac{1}{2}$ and $L_{z}-1=L-\frac{1}{2}$.
Later on we shall denote by $L_{z}$ the thickness of the
system for which we perform the  computations  and by the  half-integer quantity   
$L=L_{z}-\frac{1}{2}$ the variable  the critical Casimir force depends on. 

From the  general theory of  finite-size scaling ~\cite{fisher_nakanishi,FSS} and based on
renormalization-group analyses~\cite{krech:92}
we expect that in the scaling limit the Casimir force  takes  the universal scaling form   
\be
\label{eq:scf}
f_\Cas(\beta,H_{1}^+,H_{1}^-,L)
=L^{-d}\vartheta\left( (L/\xi_b)^{1/\nu}{\rm sign}(t),h_{1}^+,h_{1}^-
\right)
\ee
where the scaling function $\vartheta(\tau=(L/\xi_0^+)^{1/\nu}t,h_1^+,h_1^-)$ depends on the 
spatial dimension $D$
and on  the boundary conditions on  the top and bottom  surfaces. Here 
 $\xi_b=\xi_{0}^{\pm} |t|^{-\nu}$ is the  bulk
correlation length which controls the spatial 
exponential decay of the two-point correlation function;
$\xi_0^\pm$ are  nonuniversal amplitudes above $(+)$ and below $(-)$ the bulk critical temperature $T_c$.
In the whole range of temperatures, we plot the scaling functions 
using the value $\xi^+_0=0.501(2)$~\cite{RZW} which 
is the amplitude of the 
 second moment correlation length $\xi_{\mathrm{2^{nd}}}$;
for the
Ising  model
$\xi_b/\xi_{\mathrm{2^{nd}}} \simeq 1$ for $\beta < \beta_c$ ~\cite{PV}.

Below we shall  use the following notations:
$\vartheta^{(I)}(\tau, h_1)=\vartheta(\tau, h_1^+=\infty,h_1^-)$, $\vartheta^{(II)}(\tau, h_1)=\vartheta(\tau, h_1^+=|h_1^-|,h_1^-)$,
and $\vartheta^{(III)}(\tau, h_1)=\vartheta(\tau, h_1^+=0,h_1^-)$.


\section{Numerical method}
\label{sec:method}
We compute the 
free energy difference $\Delta \hat F(\beta,H_{1},L,A)$ by
using the so-called 
 coupling parameter approach  (see, e.g., Refs.~\cite{Mon} and \cite{PRE}). 
This is a viable alternative to the method
used in Ref.~\cite{DK}, 
in which a suitable lattice stress tensor has been introduced in such a way 
that its ensemble average renders $\Delta \hat F$. So far, this 
latter method can be implemented  only  for periodic BC.
%
%
\begin{figure}
\includegraphics[width=0.5\textwidth]{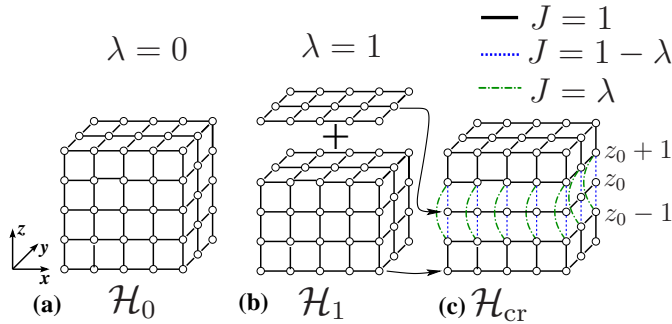}
\caption{
Bond arrangement for the computation of the free energy 
difference in Eq.~\protect{\reff{DF}} between systems of thickness $L_{z}$ (a) and $L_{z}-1$ (b)
 (see the main text). The crossover
Hamiltonian ${\cal H}_{cr}$ (c) belongs to a system which interpolates 
between those
described by the Hamiltonian ${\cal H}_0$ (a) for the system of thickness
$L_{z}$ (for $\lambda=0$)
 and by the Hamiltonian  ${\cal H}_1$ (b) for the system of thickness $L_{z}-1$ plus a 2D  layer of 
 area $A$ (for $\lambda=1$). 
}
\label{fig:fig2}
\end{figure}
%

%
The coupling parameter  approach is used in order to compute 
the difference $F_1-F_0$ between  free
energies  
$F_i = -\frac{1}{\beta}\ln \sum_{\C} \exp(-\beta {\cal H}_i), \;\;i=0,1$, 
 of models  characterized by two different energies as given by Hamiltonian  ${\cal H}_{0}$ and ${\cal H}_{1}$.
Such a calculation is successful if  the  configuration space $\C$ (i.e., the whole set of spins)
is the same for both models.
In order to implement this approach, one introduces an interpolating 
system   with the  crossover 
Hamiltonian
\be
{\cal H}_{\rm cr}(\lambda)=(1-\lambda){\cal H}_0 +\lambda {\cal H}_1.
\label{Hcr}
\ee
As a function of the coupling
parameter $\lambda\in \left[ 0,1\right ]$, ${\cal H}_{\rm cr}(\lambda)$ 
interpolates between ${\cal H}_0$ and ${\cal H}_1$ as $\lambda$
increases from 0 to 1. Accordingly the free energy 
$F_{\rm cr}(\lambda) =  -\frac{1}{\beta}\ln \sum_{\C} \exp(-\beta
{\cal H}_{\rm cr}(\lambda))$ of the crossover system
interpolates between $F_0$ and $F_1$. 
The sum is taken over all spin configurations $\C$ of the model, which
are the same for $F_{0},\;F_{1}$, and $F_{\rm cr}$. 
The difference $F_1-F_0$ can  trivially be expressed as
$F_1-F_0 = \int_0^1  F'_{\rm cr}(\lambda) {\rm d} \lambda$ where
$F'_{\rm cr}$ is the derivative of $F_{\rm cr}(\lambda)$ with respect to the
coupling parameter:
\be
\frac{\rmd F_{\rm cr}(\lambda)}{\rmd \lambda} =
\frac{\sum_\C ({\cal H}_1-{\cal H}_0)
 \rme^{-\beta {\cal H}_{\rm cr}(\lambda)}}{\sum_\C 
 \rme^{-\beta {\cal H}_{\rm cr}(\lambda)} }
=\la\Delta {\cal H}\ra_{\rm cr}(\lambda)\,,
\ee
which takes the form of the canonical ensemble average $\la \ldots
\ra_{\rm cr}(\lambda)$ 
of the energy difference
$\Delta {\cal H} \equiv {\cal H}_1 - {\cal H}_0$ 
with respect to the crossover Hamiltonian
${\cal H}_{\rm cr}$ for a given value of the coupling parameter $\lambda$.
The energy difference $\la\Delta {\cal H}\ra_{\rm cr}(\lambda)$
 can be
computed efficiently via MC simulations of the lattice model characterized by the Hamiltonian 
${\cal H}_{\rm cr}$.
Finally,  the difference of free energies is expressed 
as an integral
over the mean energy difference (see, e.g., Ref.~\cite{Mon}):
\be
F_{1} - F_{0} = \int_{0}^{1} \la\Delta
{\cal H}\ra_{\rm cr}(\lambda)\,{\rm d} \lambda.
\label{DF}
\ee

According to Eq.~\reff{eq:force} we are interested in 
the difference $\Delta \hat F(\beta,H_{1},L,A)$ 
between the free energies $\hat F(\beta,H_{1},L_{z},A)$ and $\hat F(\beta,H_{1},L_{z}-1,A)$
(we recall that  $L=L_{z}-\frac{1}{2}$ ).
In order to apply the method described above for the computation of
$\Delta \hat F(\beta,H_{1},L,A)$ (which renders $\hat f_{\Cas}$ (see Eq.~(\ref{eq:force}))) one identifies 
the model,  the Hamiltonian ${\cal H}_0$, and the associated 
configuration space $\C$ with the corresponding quantities of
the model we are interested in on the lattice $A\times L_{z}$
(see  Fig.~\ref{fig:fig2}(a)) so that 
$\hat F_0(\beta,H_{1},L_{z},A) = \hat F(\beta,H_{1},L_{z},A)$.
The final system ${\cal H}_1$ is 
identified with the slab of  area $A$ and thickness  $L_{z}-1$
plus a two-dimensional layer of size $A$:
 $\hat F_1(\beta,H_{1},L_{z},A) = 
 \hat F(\beta,H_{1},L_{z}-1,A) + \hat F_{2D}(\beta,A)$
 (see Fig.~\ref{fig:fig2}(b)).
Here $\hat F_{2D}(\beta,A)$ is the free energy 
of the isolated 2D layer of area $A$.
One has to include this 2D layer into the consideration in order
to maintain  the same number of spins in the  
configuration space $\C$ for the initial, intermediate,  and final models.
This layer can be extracted from the initial  model  at any position
$z_{0}=1,2,\ldots,L_{z}$  along the $z$-direction.
It decouples from the rest of the lattice 
upon passing from $\lambda=0$ to
$\lambda=1$, i.e., from Fig.~\ref{fig:fig2}~(a) to (b) via (c). 
The corresponding crossover Hamiltonian 
${\cal H}_{\rm cr}(\lambda)$ (but not the result of the integration in Eq.(\ref{DF}))
does depend on the position  $z_0$  from where the $2D$ layer is extracted.
In our simulations we use $z_{0} = L_{z}/2$ for even values of $L_{z}$
and $z_{0} = (L_{z}-1)/2$ for odd values of $L_{z}$. 
The explicit expression for the energy difference ${\cal H}_1-{\cal H}_0$ is
\begin{eqnarray}
\Delta {\cal H} & = & -\sum_{  x,y }\left( s_{x,y,z_0-1}  s_{x,y,z_0+1}\right. \\ \nonumber
 &- & \left.  s_{x,y,z_0-1}  s_{x,y,z_0}-  s_{x,y,z_0} 
 s_{x,y,z_0+1}\right) \;,
\label{eq:deltaH}
\end{eqnarray}
where the three indices $ (x,y,z)$ identify a lattice site,
the sum is taken over all lateral lattice site positions  in  the $xy$ plane,
and with a coupling strength $J=1$ (indicated by solid bonds in
Figs.~\ref{fig:fig2} (a) and (b); $J$ is absorbed into $\beta$). 
The crossover Hamiltonian 
${\cal H}_{\rm cr}(\lambda) = {\cal H}_0 + \lambda \Delta {\cal H}$ 
is characterized by the coupling
constants depicted in Fig.~\ref{fig:fig2}(c).
The free energy difference 
$\Delta \hat F$ (see
Eqs.~\reff{eq:force} and~\reff{DF}) can be  expressed as
\be
\Delta \hat F(\beta,H_{1},L,A) = -  \int_{0}^{1}  \langle \Delta 
{\cal H} \rangle_{\rm cr}(\lambda) {\rm d}\lambda +  F_{2D}(\beta,A)
\label{eq:df}
\ee
where the integral is taken for fixed values of $\beta$
and $H_{1}$. Note that although $\Delta {\cal H}$ is independent of $H_1$, the dependence  of
$ \Delta \hat F$ on $H_1$ enters via the statistical weight $\sim \exp(-\beta {\cal H}_{\rm cr})$.
The free energy  $F_{2D}(\beta,A)$ of the 2D layer 
can be computed from the analytical
expressions given in Ref.~\cite{Kauf}.

Once $\Delta \hat F(\beta,H_{1},L,A)$ has been computed, one has still to subtract 
$ f^\bulk(\beta)$ from it (see Eq.~\reff{eq:force}) in order to obtain the
Casimir force for a slab of assigned thickness $L=L_{z}-1/2$.
We determine the bulk free energy  by using the  temperature integration 
method~\cite{hucht,hasen1,hasen2} applied to 
 a cubical system of size $L_{\rm cube}$
with periodic boundary conditions.
For such a system the free energy per site (in units of  $k_BT$) can be written  as 
\be
 \beta\hat f(\beta,L_{\rm cube})=-\ln(2)+\frac{1}{L_{\rm cube}^{3}}
\int \limits_{0}^{\beta} \la E (\beta',L_{\rm cube}) \ra {\rm d }\beta',
\label{eq:fbulk}
\ee
where $ \la E (\beta',L_{\rm cube})\ra$ 
is the averaged internal energy of the system 
at the inverse temperature $\beta'$ and for the size $L_{\rm cube}$; 
 $-\ln(2)$ is the free energy in units of $k_BT$ and per site at  $\beta=0$.
For a cube, in the limit $L_{cube}\to \infty$ the finite-size dependence of the 
free energy density for a cube 
is predicted \cite{FSS} to  scale with $L_{cube}$ as
$\beta \hat  f(\beta,L_{\rm cube})- \beta f^{\rm bulk }(\beta) \propto L^{-3}_{\rm cube}$.
Therefore the bulk free energy per spin follows as the limit  $\beta f^{\rm bulk }(\beta)=
\lim \limits_{L_{\rm cube} \to \infty} \left[ \beta \hat f(\beta,L_{\rm cube})\right]$.
At the critical point    one has 
\be
\beta_c \hat f(\beta_{c},L_{\rm cube})\simeq
\beta_c f^{\rm bulk }(\beta_{c})+U_0L^{-3}_{\rm cube},
\label{eq:ccorr}
\ee
with  the universal finite-size scaling amplitude $U_0 = -0.657(3)$ (see  Ref.~\cite{Mon}).

In order to determine the universal  scaling function of the critical Casimir
force we perform the following steps (details are given below).
For each temperature we compute the averaged
internal energy $\la E (\beta,L_{\rm cube})\ra $ for  a cube with periodic boundary conditions
by using a histogram reweighting MC method.
Then we carry out a numerical integration in order 
to obtain   an estimate for the bulk free energy 
$\beta f^{\rm bulk}(\beta)$ in accordance  with Eqs.~(\ref{eq:fbulk}) and 
(\ref{eq:ccorr}).  
For the slab geometry $ A \times L_{z}$, at the inverse temperature $\beta$,
 and for a fixed  boundary field $H_{1}$,
  we compute 
the ensemble averages $\la \Delta {\cal H}\ra_{ cr}(\lambda)$ via MC simulations for
 $N_{\lambda}=21$
 different values of $\lambda_{k}=\frac{k}{N_{\lambda}-1}, k=0,\ldots, N_{\lambda-1}$. 
Based on these $N_{\lambda}$ values  we carry out the  numerical integration in Eq.~(\ref{eq:df}) and use
  an analytical expression for $F_{2D}(\beta,A)$ in order  to
obtain $\Delta \hat F(\beta,H_{1},L,A)$.  
Combining the results for the bulk free energy density
$\beta f^{\rm bulk }(\beta)$  and for the free energy difference
and by using  Eq.~(\ref{eq:force})
we obtain a  numerical estimate
for the critical Casimir force $\hat f_\Cas(\beta,H_{1},L,A)$. 
In order to obtain the corresponding 
scaling function $\hat \vartheta$ we perform computations for various values of $L$, $A$, the inverse temperature $\beta$,
and boundary fields $H_1$.
 The scaling function $\hat \vartheta$ in
Eq.~\reff{eq:scf} is retrieved from the numerical data for $ \hat
f_\Cas$ by taking into account finite-size corrections as described in 
the following section.

For determining the bulk free energy density  the 
histogram reweighting method has been  used as follows ~\cite{FS,LB}.
The computation of the energy distribution 
$P(E,\beta_{i})$ has been performed for a choice of 256 points  $\beta_{i} \in [0,0.3]$ 
for a cubic system of size $L_{\rm cube}=128$. 
For the numerical simulation we have employed the
hybrid MC method,  which is a suitable mixture of Wolff  
and  Metropolis algorithms  ~\cite{LB}.
For thermalization $4 \times 10^{5}$ hybrid MC steps have been  used.
The averaging has been performed  over $10^{6}$  hybrid MC steps which 
have been  split into 10 series for the evaluation of statistical errors. 
Therefore,
 for every value $\beta_{i}$ actually ten histograms (each consisting of 
$10^{5}$ MC steps) have been  computed.
According to the histogram reweighting method 
 one can  obtain an estimate for $\la E \ra$ at 
 an inverse temperature $\beta'$
based on the histogram $P(E,\beta_{i})$ 
for the inverse temperature $\beta_{i}$  ~\cite{FS,LB}:
\be
\la E \ra_{\beta_{i}}(\beta')
=\frac{\sum \limits_{ E } E P(E,\beta_{i})e^{-E(\beta'-\beta_{i})}}
{\sum \limits_{ E } P(E,\beta_{i})e^{-E(\beta'-\beta_{i})} }.
\ee
For every    $ \beta' \in [\beta_{i}, \beta_{i+1}]$
we define the interpolated internal energy
\be
\la E \ra(\beta')=
\frac{\beta_{i+1}-\beta'}{\beta_{i+1}-\beta_{i}} \la E \ra_{\beta_{i}}(\beta')
+
\frac{\beta'-\beta_{i}}{\beta_{i+1}-\beta_{i}} 
\la E \ra_{\beta_{i+1}}(\beta').
\ee
We have checked that for the same inverse temperature $\beta'$ the difference between 
the estimates   $\la E \ra_{\beta_{i}}(\beta')$
and $\la E \ra_{\beta_{i+1}}(\beta')$,
 which use histograms
for two neighboring points $\beta_{i}$ and $\beta_{i+1}$,
is substantially less than the statistical inaccuracy
of our simulation data. The  statistical inaccuracy has been
 determined canonically 
over 10 series of histograms.
In the next step, in accordance with Eq.~(\ref{eq:fbulk}) we obtain the free energy 
$\beta\hat f(\beta,L_{cube})$ 
by integrating numerically the interpolated internal energy.
For the intergration we  employ the trapezoidal rule  with a large  $(>10^5)$ 
number of points, so that the  inaccuracy
of the numerical integration is less then $10^{-9}$.
We estimate that at the bulk critical temperature  $\beta_{c}$
 the statistical error $\Delta \beta_c\hat f(\beta_c,L_{cube})$ 
for the 
free energy determined from 10 series  is  about $4 \times 10^{-7}$.
In the following,
 we neglect the finite-size correction of the bulk free
 energy and take $\beta f^{\rm bulk }(\beta)\simeq
  \beta \hat f(\beta,L_{\rm cube}=128)$  (compare  Eq.~(\ref{eq:ccorr})).
This is justified, because
for the maximal value $L=19.5$  used in our simulation the finite-size correction to $\vartheta$
 due to the finite system size $L_{\rm cube }=128$, i.e.,   
$0.657(L/L_{cube})^3 \simeq 0.0023$,
is of the same order  as the statistical error stemming from 
the contributon 
 $L^3\Delta \beta_c\hat f(\beta_c,L_{cube})$, 
i.e., $(19.5)^3\times 4 \times 10^{-7}\simeq 0.00297$ 
(see Eqs. (\ref{eq:force}) and (\ref{eq:scf})).

\begin{figure}[h]
\includegraphics[width=0.45\textwidth]{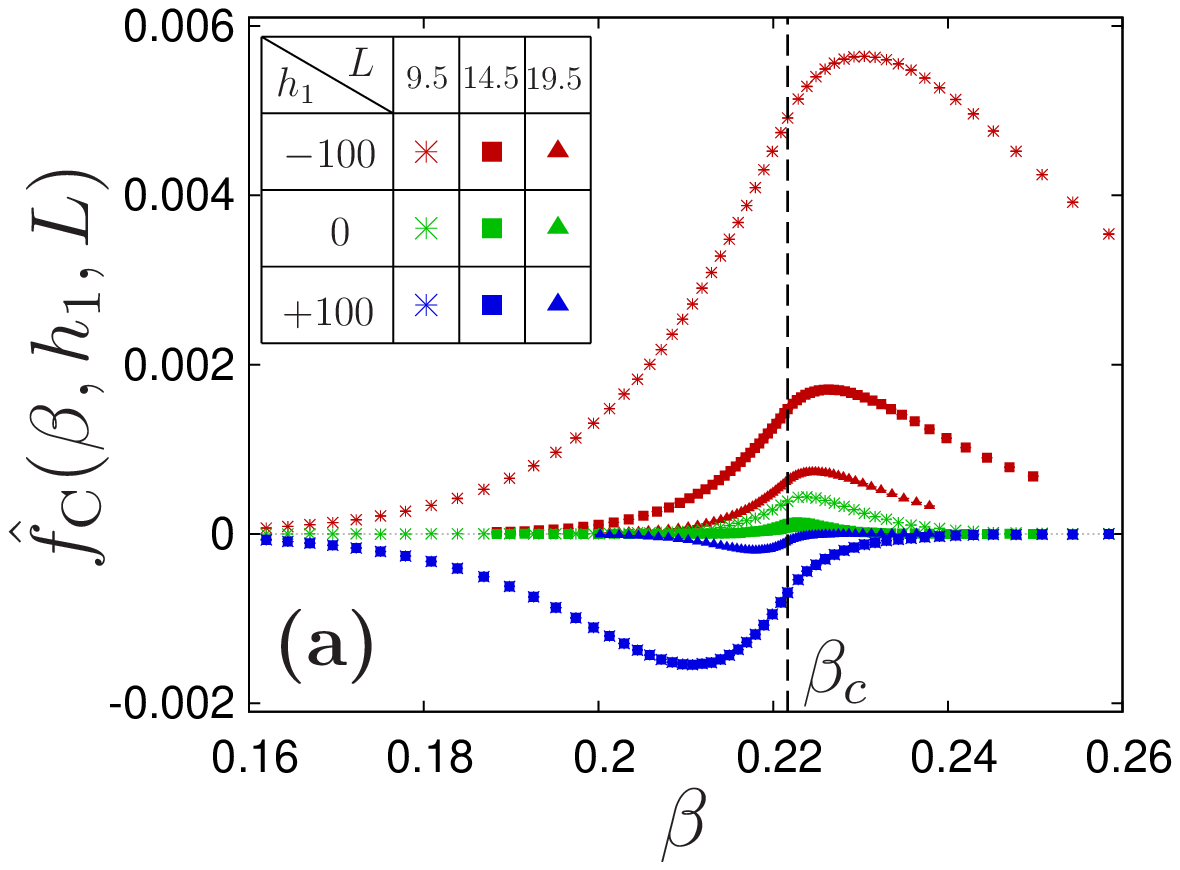}

\includegraphics[width=0.45\textwidth]{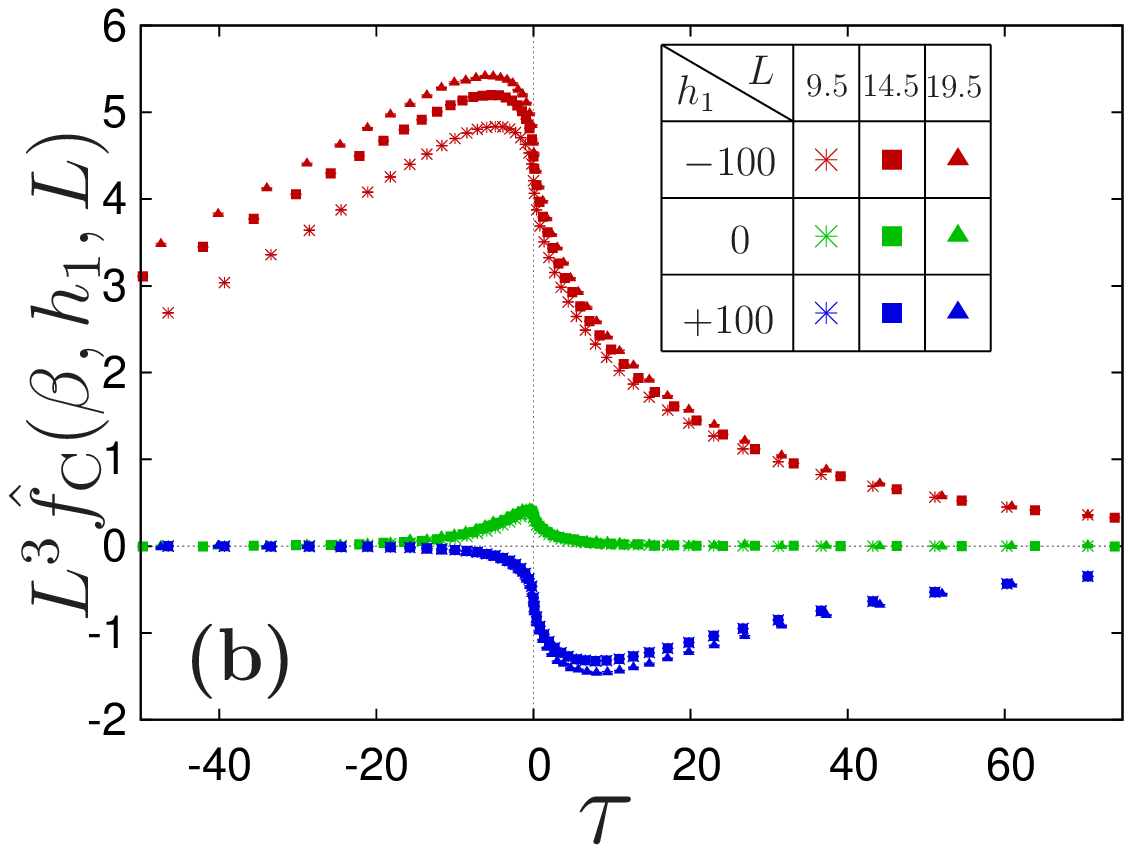}
\caption{ MC data for case (I):
 (a) Casimir force ${\hat f}_{\Cas}$  as a function of the inverse
temperature $\beta$; $\beta_c=0.2216544(3)$
(b) rescaled 
 Casimir force $L^{3}{\hat f}_{\Cas}$ as a function of the 
scaling variable  $\tau=(L/\xi_0^+)^{1/\nu}(T-T_{c})/T_{c}=(L/\xi_0^+)^{1/\nu}t$.
The data correspond to  $ h_{1}=-100,0,100$ and $ L= 9.5, 14.5, 19.5$. In (a) and (b) the data
for $(h_1,L)=(100,9.5)$ and $(100,14.5)$ as well as $(0,14.5)$ and $(0,19.5)$ can be barely distinguished.
}
\label{fig:fc}
\end{figure}

For the computation of the free energy $\Delta \hat F(\beta,H_{1},L,A)$ 
in Eq.~(\ref{eq:force})
we use  slabs of thicknesses $L_{z}=10,15,20$,
so that $L=9.5,14.5,19.5$,
with an aspect ratio equal to 6: $L_{x}=L_{y}=6L_{z}$, $A=36L_{z}^{2}$.
In order to compute  the  average 
$\la \Delta {\cal H}\ra_{\rm cr}(\lambda)$ we again use the
hybrid MC method  with a mixture of Wolff 
and  Metropolis algorithms. Each hybrid MC step consists of a
 flip of a Wolff cluster according to the Wolff algorithm, 
 followed first by $3A$  attempts
to flip an arbitrary spin and then  by $3A$
attempts to flip a spin $s_{x,y,z}$ with $z\in \{z_0-1,z_0,z_0+1\}$. 
These attempts are accepted according to the Metropolis rate~\cite{LB}.
We use $2.5 \times 10^{5}$ MC steps for thermalization.
For the computation of the thermal 
average we use $5 \times 10^{5}$ MC steps 
 split into  10 series.
For each  series, using Simpson's rule we perform a numerical integration over $N_{\lambda}=21$
points for fixed values of the inverse temperature $\beta$, the surface field $H_1$,
 and the width $L$ of the slab.
Having computed  the free energy difference $\Delta \hat F(\beta,H_{1},L,A)$, 
 for each series we  
finally  combine the results for the bulk free energy 
$\beta \hat f^{\rm bulk}(\beta)$
with the corresponding ones  for the free energy difference
 $\Delta \hat F(\beta,H_{1},L,A)$ and determine the numerical inaccuracy.

In Fig.~\ref{fig:fc}(a) we plot the Casimir force ${\hat f}_{\C}$
as a function of $\beta$ for the three values $h_1=-100,0,100$
of the bottom surface scaling  field.
In Fig.~\ref{fig:fc}(b)
we plot the rescaled values of the Casimir force $L^{3}{\hat f}_{C}$
as a function of the temperature scaling variable 
$\tau=(L/\xi_0^+)^{1/\nu}(T-T_{c})/T_{c}=(L/\xi_0^+)^{1/\nu}t$ for  case (I). The 
visible absence of the expected
data collapse is due to  finite size corrections
to scaling, which will be discussed in the following section.

\section{Finite size corrections to scaling}
\label{sec:corr}

Finite-size scaling  is known to be  valid asymptotically for finite but 
large lattices  and small values of $t$, i.e., for  a large bulk correlation length
$\xi_b$; here large means relative to the lattice constant  ~\cite{FSS}. Outside
the asymptotic regime corrections to the leading (universal) scaling behavior
become relevant. These non-universal 
corrections affect both the scaling variables
and the scaling functions and depend on the details of the model
as well as on the geometry and the boundary conditions~\cite{privman,luck}.
Renormalization-group analyses reveal that 
there is a whole variety of sources for corrections to scaling  which arise
from bulk, surface, and finite-size effects~\cite{FSS}. 
%
%
%
\begin{figure}[h]
\includegraphics[width=0.45\textwidth]{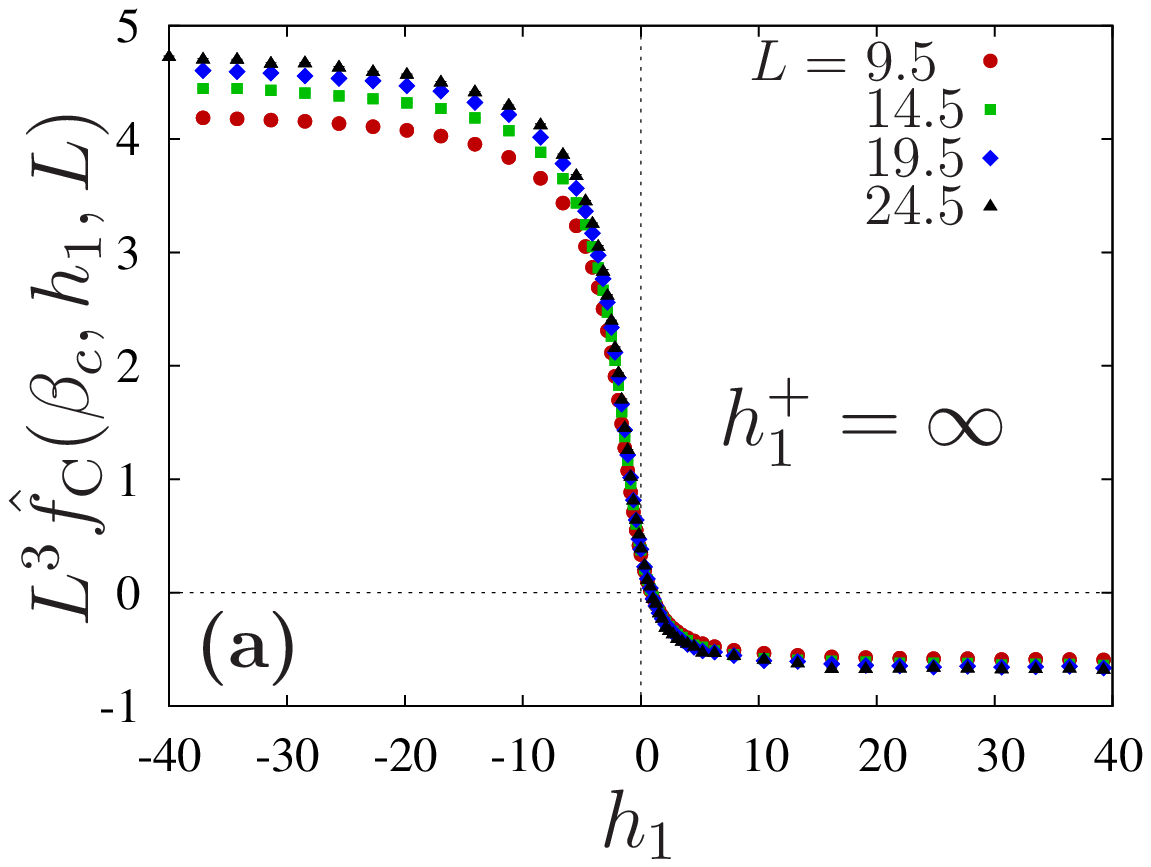}

\includegraphics[width=0.45\textwidth]{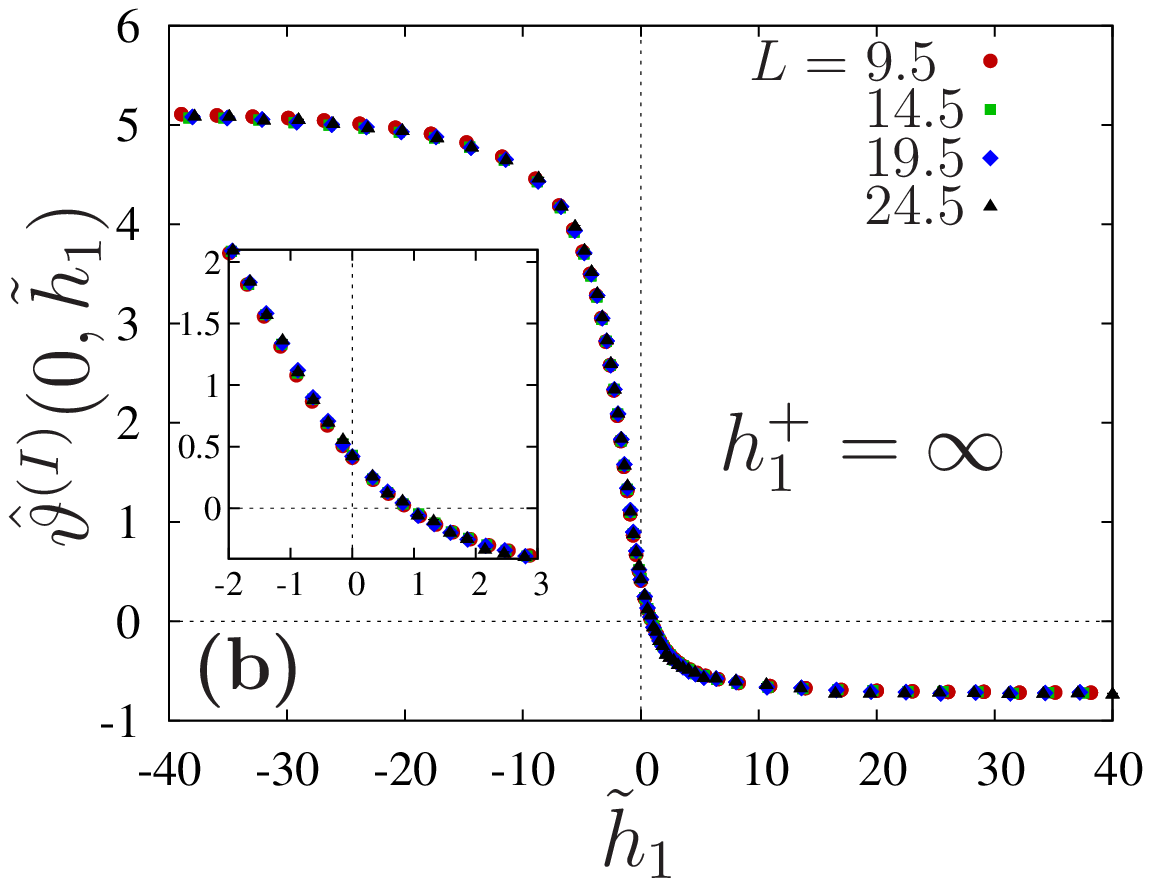}
\caption{
Results at the critical point $\beta=\beta_c$ for  case
 (I)  (see Fig.~\ref{fig:fig1}) as a function of boundary fields:
(a) Casimir force multiplied by $L^{3}$ as a function of the scaling
variable $h_{1}=H_{1}L^{\Delta^{ord}_{1}/\nu}$ without finite size corrections;
(b) Casimir force scaling function $\hat \vartheta^{(I)}(\tilde \tau=0,\tilde h_1)$ 
with taking corrections $L+\delta$ into account
as a function of the corrected scaling variable 
$ \tilde h_{1}=H_{1}(L+\delta)^{\Delta_{1}^{ord}/\nu}$; in units of the lattice spacing  $\delta=0.65$
at the bulk critical point $\beta_{c}$. For $\tilde h_1=0$ one finds the fixed-point $(O,+)$.
}
\label{fig:fchi}
\end{figure}
%
%

%
For the finite and rather limited sizes of the lattices which  we 
investigate in  our MC simulations, 
it is necessary to take  corrections to scaling  into account in order to
obtain data collapse and thus allowing us to infer the leading universal scaling function
~\cite{hucht,EPL,PRE}.

In the present study, the following quantities are expected to acquire
corrections to scaling:
\begin{itemize}
\item  the amplitude of the scaling function $\vartheta =L^{3} f_{\C}$
\item  the surface field scaling variable $ h_1$
\item  the temperature scaling variable $\tau=t\left(L/\xi_{0}^{+} \right)^{1/\nu}$.
\end{itemize}
In our previous MC simulations aimed at obtaining 
 critical Casimir forces  for Ising  films
with a variety of {\it universal}  boundary conditions, such as $(+,+), (+,-)$, 
or $(O,O)$ BC  ~\cite{EPL,PRE}, corrections to scaling were taken into account 
by using various ans\"atze. The choice  for a particular form
of corrections to scaling was guided by achieving 
the best  data collapse or the best fits used in our computations.
For example,  for the amplitude of the scaling function we adopted the expression
$
\hat f_{\Cas}=
L^{-3}\frac{(1+ g_{1}L^{-1})}{(1+ g_{2}L^{-1})}\sfh
$. Various  variants for this form of corrections to scaling were considered;
 we used $(g_1 \ne 0,g_2), (g_1,g_2\ne 0)$, or $(g_1\ne 0,g_2\ne 0)$.  They all lead to a satisfactory
data collapse, but the inferred amplitude of the scaling function of the critical Casimir force depends sensitively
on the particular ansatz.

In Refs.~\cite{hasen1,hasen2}  still another type
 of finite-size correction
is employed. It amounts to introducing an effective width  $ L+\delta $ of the slab so that, 
e.g., the amplitude of the scaling function of the critical Casimir force scales as 
$\hat f_{\Cas}= 
(L+\delta)^{-3}\sfh$. Using this type of finite-size correction
may be justified as follows.
As mentioned in  Sec.~\ref{sec:intr} surfaces  subjected 
to the action of a  surface field asymptotically belong 
to the  surface universality class of the normal transition which 
corresponds to $(+)$ or $(-)$
fixed-point  boundary conditions in the sense of 
  renormalization-group theory \cite{binder:83:0,diehl:86:0}.
(The  $(+)$  and $(-)$ boundary conditions are realized as the limits 
of the scaling field   $h_1^{(i)}
\rightarrow +\infty$ and $-\infty$, respectively.)
For such boundary conditions, on the
coarsed-grained scale the
 order parameter varies as 
$|\phi(z\to 0)| \propto z^{-\beta/\nu}$  for small normal distances from the surface
(but still large on molecular scales) \cite{diehl:97,leibler}. Within a certain range of small $z$ values
such a divergent
 behavior is expected to hold also 
for a finite but sufficiently strong  surface field.  
In Ising   {\it lattice} models with boundary conditions
 corresponding to $(+)$ or $(-)$
fixed-point  BCs, 
 the  order parameter does not diverge at the surface but   saturates there
at the value $+1$ or $-1$. 
Changing the width of the slab from $L$ to $L +\delta$
with a nonuniversal length $\delta = z_{\rm ex}^{(1)}+z_{\rm ex}^{(2)}$ such 
that  the order
parameter profile behaves as $|\phi(z\rightarrow 0)| \sim
(z+z_{\rm ex}^{(i)})^{-\beta/\nu}$   \cite{binder:83:0,SDL-94} 
upon approaching the
wall $i$,   turns out to be an
effective means to take into account corrections to the leading critical
behavior \cite{SDL-94}; $z_{\rm ex}^{(i)}$ plays 
the role of an extrapolation length \cite{binder:83:0,diehl:86:0}.
Similarly, the effects  of a physical
wall with a {\it finite} surface field (which implies $\ell_1^{(i)} \neq 0$ 
( see Eq.~(\ref{eq:length})))
on the order parameter are equivalent to those of a fictitious wall with strong
surface fields  (which means $\ell_1^{(i)}=0$) displaced by a distance 
$-z_{\rm ex}^{(i)}$ from
the physical wall. 
One can  determine the length  $\delta$ by analyzing the spatial variation 
of the order parameter profile, as it was done for the Blume-Capel model in Ref.~\cite{hasen2}.
Here, we assume  that the equivalence  described above  carries over
to  critical Casimir forces such that we can  determine the effective 
width $L+\delta$ of the slab 
by demanding the best data collapse. 
We apply this method  also in the crossover regime, i.e.,
for sufficiently weak  surface fields for which 
 upon approaching the critical point one effectively 
observes a crossover to the boundary condition corresponding to the ordinary transition $(O)$
fixed point. As discussed in the introduction, the order parameter profiles in a film with weak surface
fields deviate strongly from the fixed-point universal behavior. Accordingly, we expect that within 
this range
of surface fields the aforementioned type of correction does not  satisfactorily capture
 the actually corrections to scaling.

In Fig.~\ref{fig:fchi}(a) 
  we plot the 
rescaled critical Casimir force $L^{3}{\hat f}_{\C}$ for  case (I).
It is  evaluated 
at the critical point 
$\beta_{c}$ and presented  as a function of
$ h_{1}$  without  finite-size corrections  taken into account. 
Apparently the
data for the rescaled force do not coincide for various  values of  $L=9.5,14.5,19.5,24.5$.
In order to obtain the expected data collapse we apply the following finite-size corrections
(here and in the following we denote scaling variables
with finite size corrections by a tilde:~$\tilde \tau$,~$\tilde h_{1}$):
\be
\label{eq:corr}
\hat f_{\C}(\beta,H_{1},L,A)=
 (L+\delta)^{-3} \hat \vartheta( \tilde \tau, \tilde h_1)
\ee
with
\be
\label{eq:corrx}
\tilde \tau \equiv \tau \left[
(L+\delta)/\xi_{0}^{+}\right]^{1/\nu} 
\left[1 + g_{\omega} (L+\delta)^{-\omega}\right]
\ee
and (see Eq.~(\ref{eq:sch}))
\be
\label{eq:corrh}
\tilde h_1=H_{1}(L+\delta)^{\Delta^{ord}_{1}/\nu},
\ee
where $\omega=0.84(4)$ is the  leading bulk correction-to-scaling
exponent \cite{PV}; the length 
$\delta$ and the  coefficient $g_{\omega}$ remain to be determined. 

The  value of the length $\delta$ is
obtained from the data for the critical Casimir force 
at the critical point (these data are presented in  Fig.~\ref{fig:fchi}(a)).
By using Eqs.~(\ref{eq:corr})  and (\ref{eq:corrh}) 
and by implementing the fitting procedure  within the interval 
$\tilde h_{1} \in [-15,15]$  with $\delta$  being the only  fit parameter,
we obtain the value $\delta=0.65$  which  minimizes   deviations
between data for different values of $L$. Including error bars we find 
$\delta=0.65(2)$ for various intervals
of $h_{1}$ or  $\delta=0.60(5)$ for different sets of $L$.
The final result
for the scaling function $\hat \vartheta(0,\tilde h_{1})$ with corrections to scaling corresponding to
$\delta=0.65$ 
is shown in Fig.~\ref{fig:fchi}(b). 
For large absolute  values of  $\tilde h_{1}$
we reproduce the data from Refs.~\cite{EPL,PRE} for the critical point with $(-,+)$
and $(+,+)$ BCs.
For small values of $|\tilde h_{1}|$ we observe the crossover between
these two regimes.
%
%
\begin{figure}[h]
\includegraphics[width=0.45\textwidth]{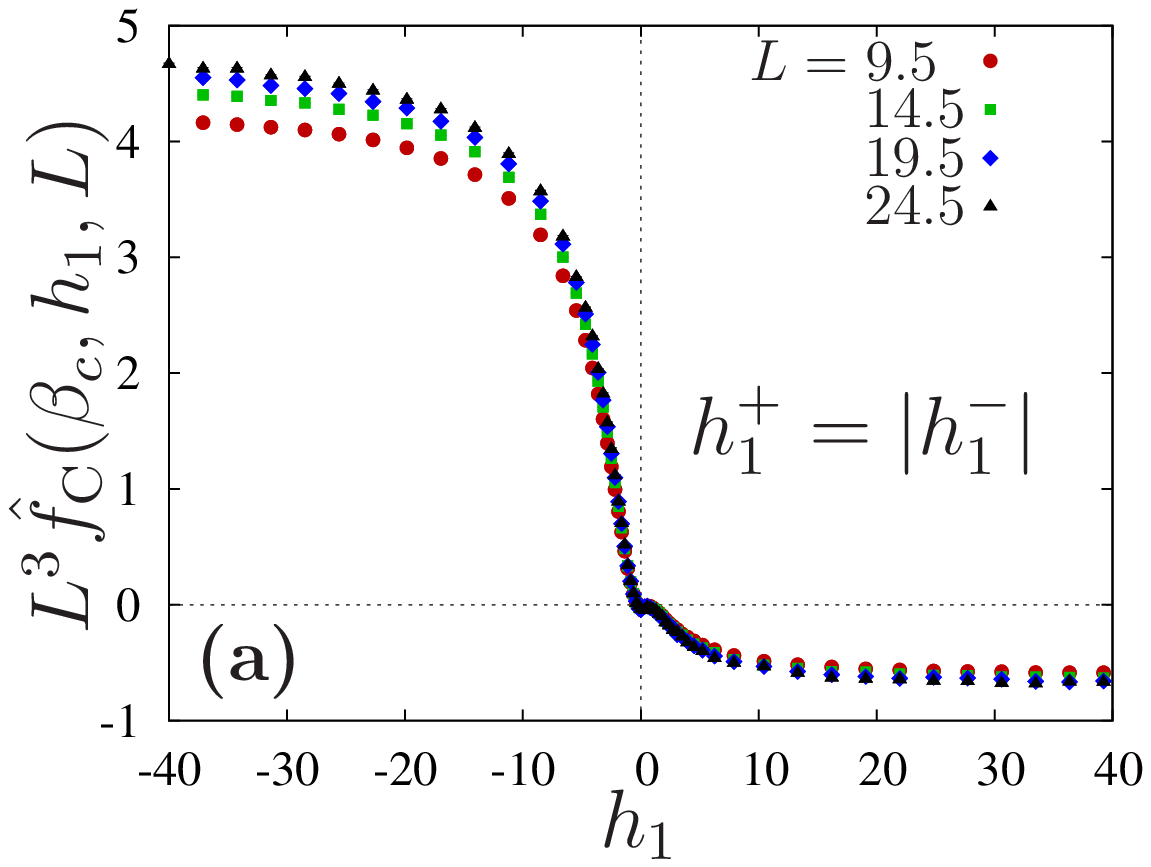}

\includegraphics[width=0.45\textwidth]{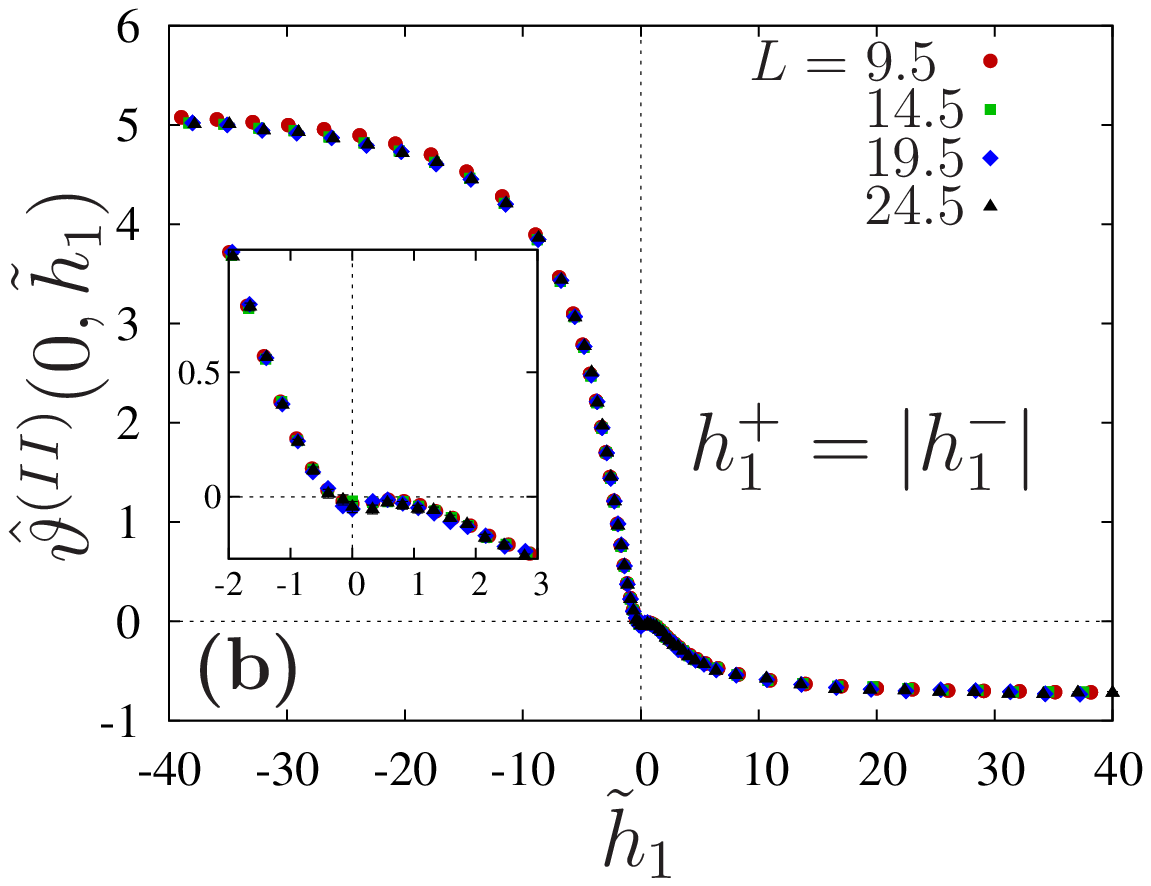}
\caption{
Same as Fig.~\ref{fig:fchi} for the case   (II) (see Fig.~\ref{fig:fig1}).
For $\tilde h_1=0$ one finds the fixed point $(O,O)$.
}
\label{fig:fchh}
\end{figure}
%
%

The  procedure which  we used in order 
to obtain the best fit for the value of the length $\delta$ 
 is described in
 detail in the appendix of  Ref.~\cite{PRE}. One of the difficulties
in finding the optimal data collapse is that the fitting function itself, i.e., the scaling function
of the critical Casimir force, is not known.
For the initial guess for the value of the length  $\delta$ 
we  infer the scaling function from the
corresponding data for ${\hat f}_{\C}$, one function $\hat \vartheta_k$ for each value $L_k$ $(k=1,\cdots, N)$ used
 (see Eqs.~(\ref{eq:corr}),
 (\ref{eq:corrx}), and (\ref{eq:corrh})).
We define an expected   scaling function as the average of the various $\hat \vartheta_k$: 
$\hat \vartheta_{expected}=(1/N)\sum_{k=1}^N \hat \vartheta_k $.
 Finally, for every $L_k$ and for a given value of $\delta$  
 we compute the sum of 
squares $\chi^{2}(\delta)$ of the deviation of the aforementioned scaling functions  
from the expected  scaling function.
Finally,  we determine as  the  value of $\delta$ the one which  minimizes  $\chi^{2}$.
That value provides the best data collapse of the  data for different $L$.

\begin{figure}[h]
\includegraphics[width=0.45\textwidth]{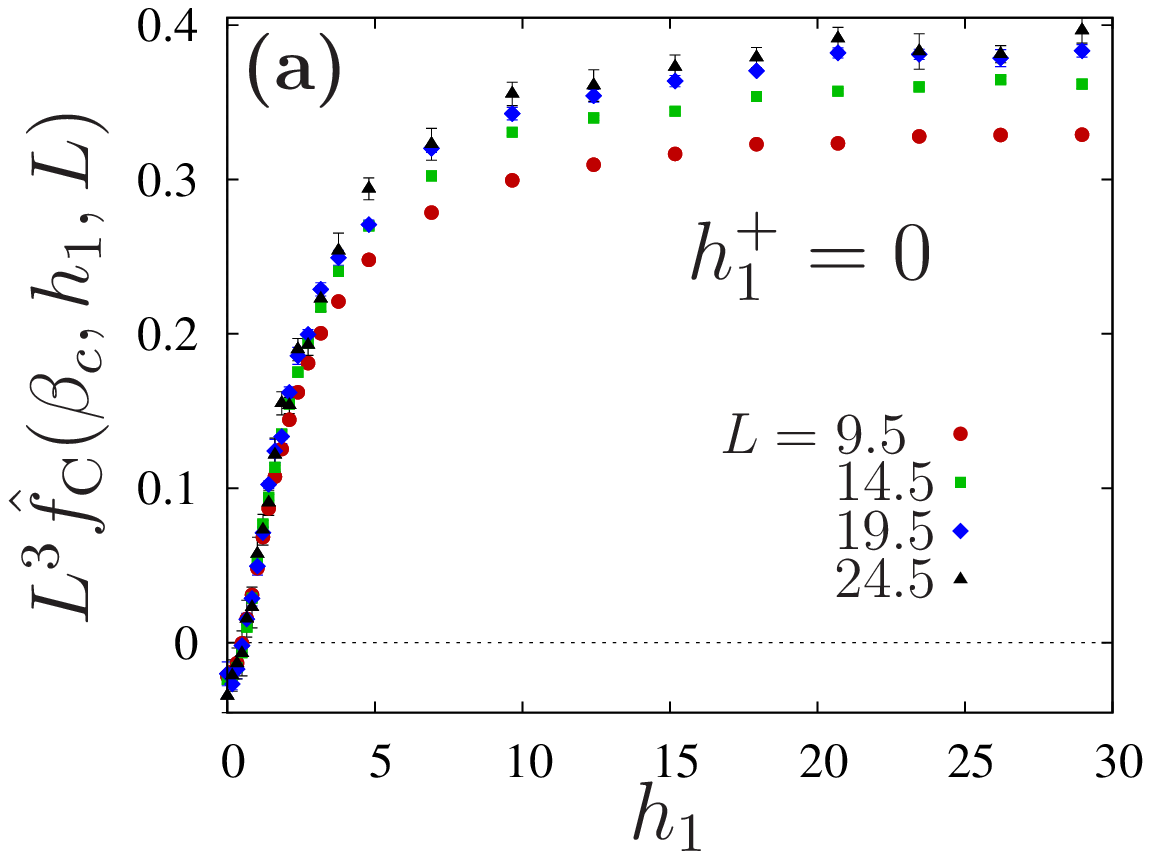}

\includegraphics[width=0.45\textwidth]{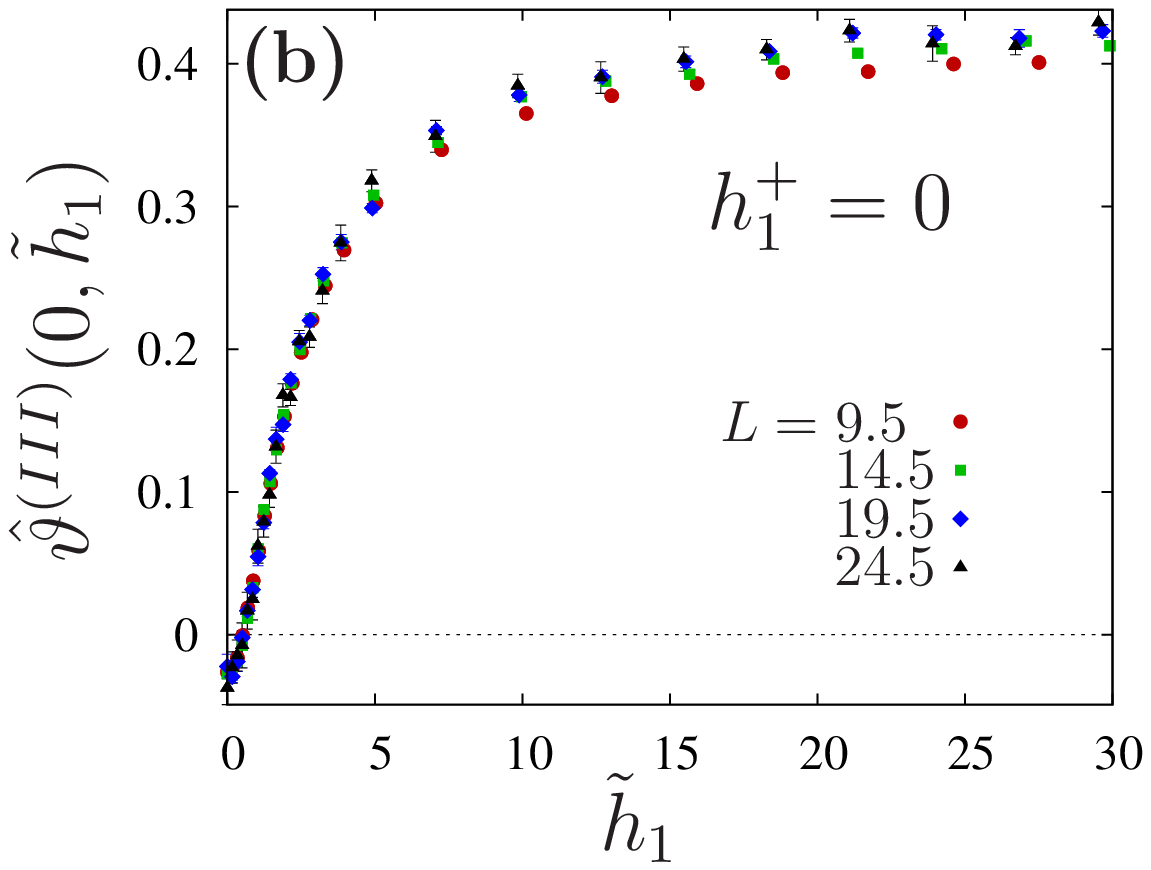}
\caption{
 Same as Fig.~\ref{fig:fchi} for the case (III) (see Fig.~\ref{fig:fig1}).
For $\tilde h_1=0$ one finds the fixed point $(O,O)$.
}
\label{fig:fcho}
\end{figure}

Applying the same procedure for case (II) and case (III)
we obtain the values $\delta=0.6$  and $\delta=1.4$, respectively.
However, in order to be consistent (as mentioned earlier, for some values of the surface field
 different cases coincide) we use the common value $\delta=0.65$
for all cases. In Figs.~\ref{fig:fchh} and \ref{fig:fcho}
we present our results  without (a) and with (b)
finite-size corrections for case (II) and case (III), respectively.
%
%
%
%
%

Knowing the finite-size corrections of the surface field scaling variable 
 we can carry out numerical
simulations for various values of  $\tilde h_{1}$;
 for each  value of $\tilde h_{1}$ we can extract information 
 about the coefficient $g_{\omega}$ using the same procedure as for the determination of
the length $\delta$.

%
\section{Results}
\label{sec:res}
\begin{figure}[!]
\includegraphics[width=0.45\textwidth]{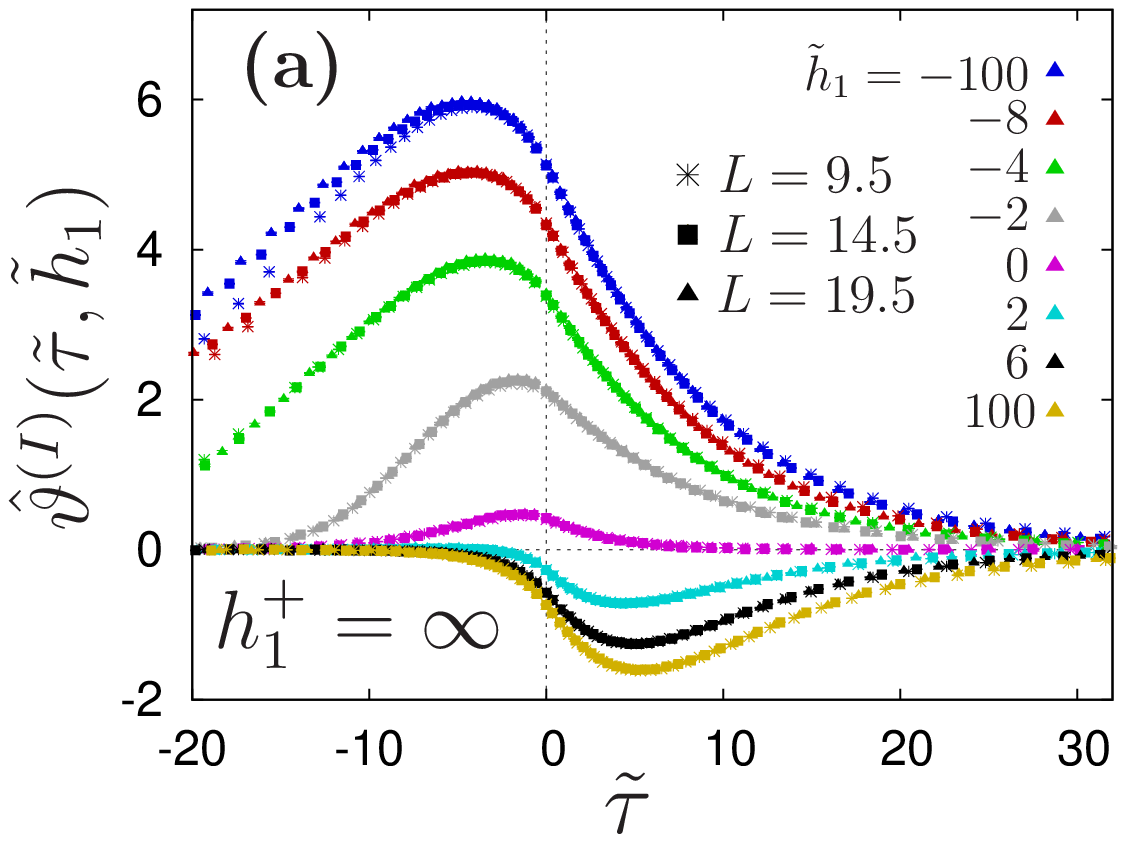}

\includegraphics[width=0.45\textwidth]{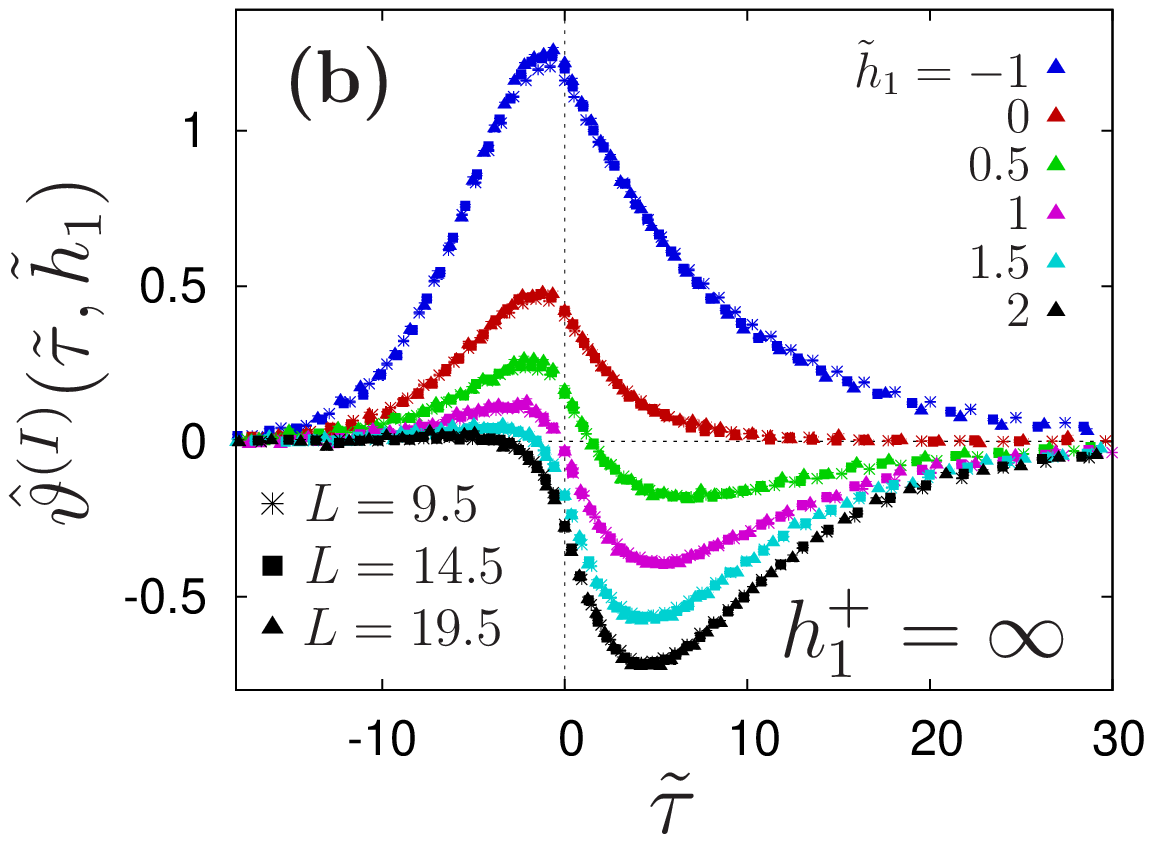}
\caption{
 The scaling function $\hat \vartheta^{(I)}$ of the critical Casimir
force  for  case (I), i.e.,  $h_{1}^{+}=\infty$
as a function of the temperature scaling variable $\tilde \tau$ (see Eq.~(\ref{eq:corrx}))
for various bottom boundary fields corresponding to certain values of the surface
field scaling variable $\tilde h_{1}\equiv \tilde h_1^-$ (see Eq.~(\ref{eq:corrh})): 
 (a) large amplitudes of the surface field (from top to bottom): 
 $\tilde h_{1}=-100,-8,-4,-2,0,2,6,100$;
(b) small amplitudes  of the surface field for which a crossover
from  repulsive to attractive forces as function of $\tilde \tau$ is observed (from top to bottom):
 $\tilde h_{1}=-1,0,0.5,1,1.5,2$.  For each color $\bigstar$ corresponds to $L=9.5$, $\blacksquare$ to $L=14.5$, and 
$\blacktriangle$ to $L=19.5$.
}
\label{fig:thetat_hi}
\end{figure}

Here we present  the critical  Casimir force scaling function determined
for various values of  $\tilde h_{1}$ 
 as a function of $\tilde \tau$.
The set of values used for $\tilde h_{1}$  is given in  
Tables~\ref{tab:I}, \ref{tab:II}, and \ref{tab:III}. 
For each of the three values  $L=9.5,14.5,19.5$  of the  slab thickness
we infer the values of the surface fields $H_{1}$ which
correspond to the pair $(\tilde h_{1},L)$ according to Eq.~(\ref{eq:corrh})
with $\delta=0.65$. Next, for each  pair $(\tilde h_{1},L)$
the  critical Casimir force $\hat f_{\C}(\beta,H_{1},L)$ has been  computed
for various inverse temperatures $\beta$.
Finally, for each value of $\tilde h_{1}$ we apply the fitting procedure described above in order 
to determine  $g_{\omega}$  by using Eq.~(\ref{eq:corrx}). Our results 
for $g_{\omega}$  are given in 
Tables~\ref{tab:I}, \ref{tab:II}, and \ref{tab:III} for the cases (I), (II), and (III),
respectively. 

\begin{figure}[!]
\includegraphics[width=0.45\textwidth]{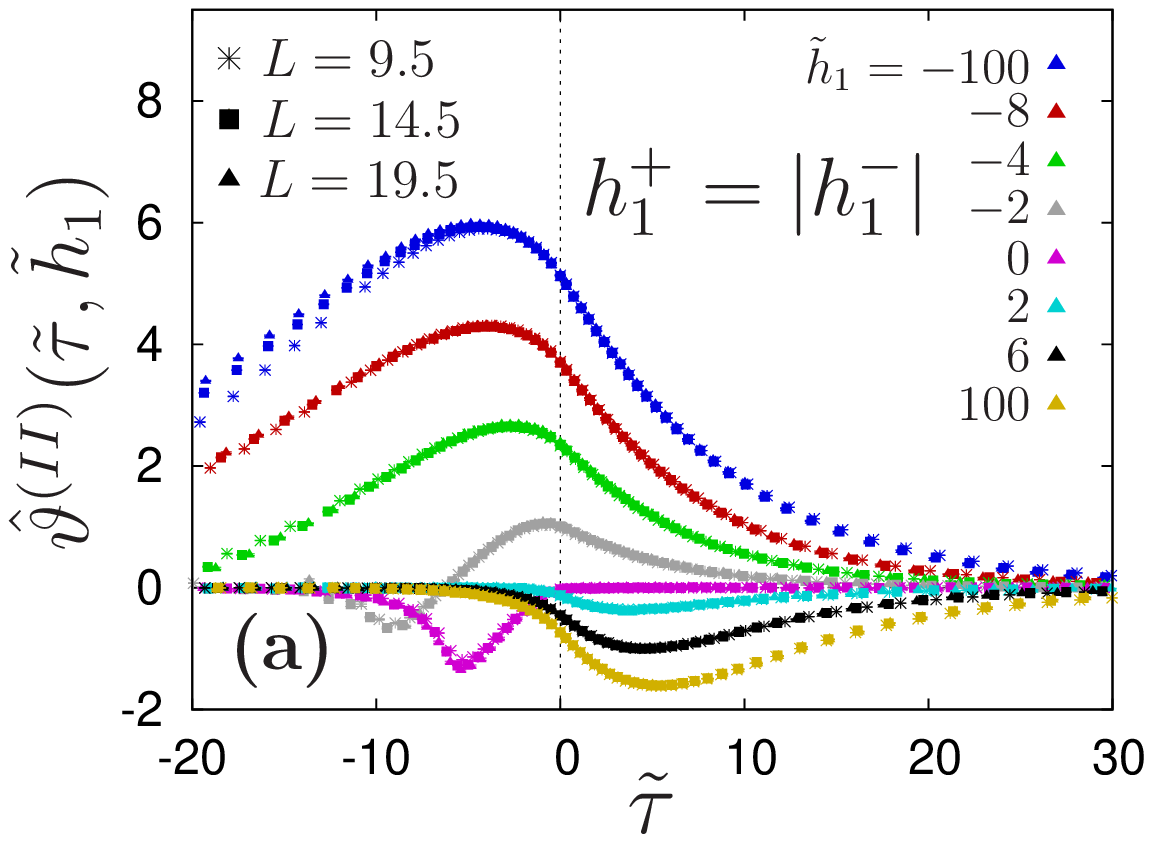}

\includegraphics[width=0.45\textwidth]{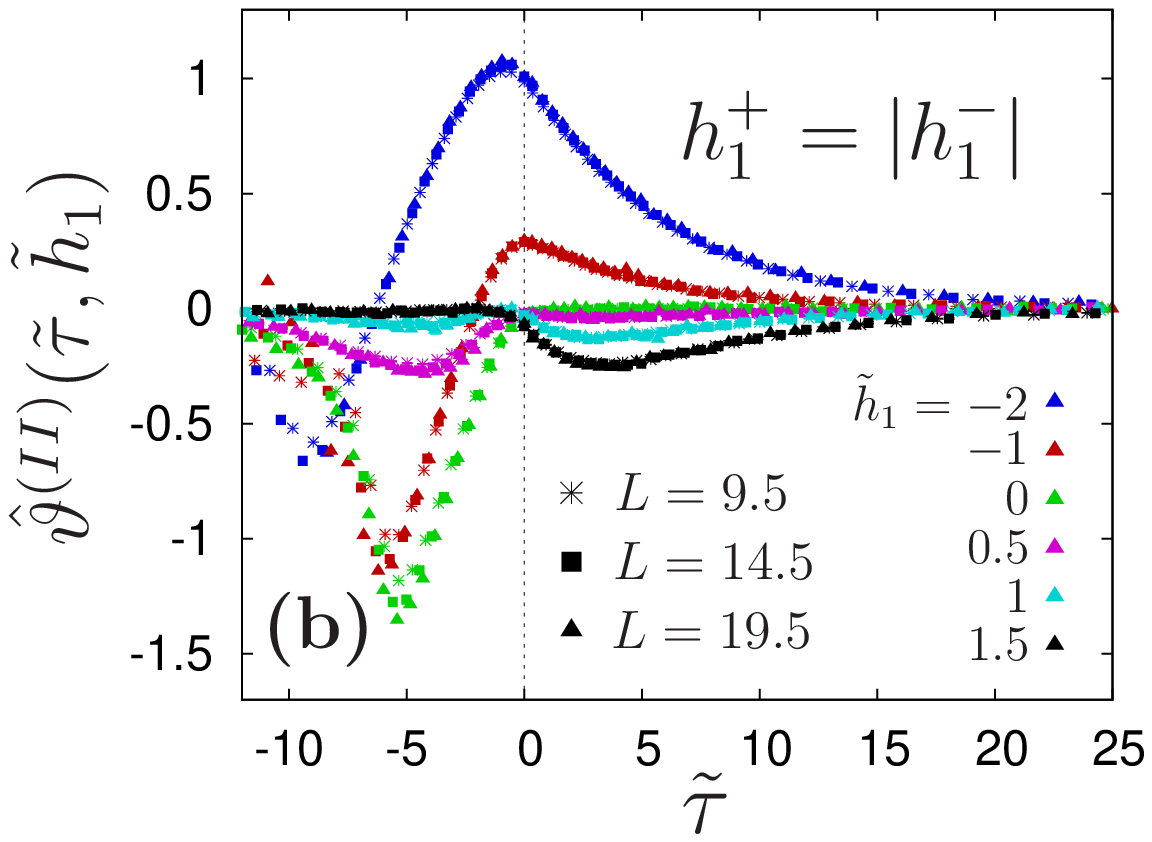}
\caption{
Same as Fig.~\ref{fig:thetat_hi} 
 for  case  (II) (see Fig.~\ref{fig:fig1}).
In (b) for small amplitudes of the surface  field  [from top to bottom:
 $\tilde h_{1}\equiv \tilde h^-_1=-2,-1,0,0.5,1,1.5$]  there occurs  a  crossover
from attractive to repulsive forces upon increasing $\tilde \tau$. }
\label{fig:thetat_hh}
\end{figure}

\begin{table}
\caption{
Values of $\tilde h_{1}$ with the  corresponding values of $g_{\omega}$ as obtained 
from the fitting procedure for case (I), i.e.,  $h_{1}^{+}=\infty$.}
\begin{tabular}{|c|c|c|c|c|c|c|}
\hline
$\tilde h_{1}$ & -100     &-8 & -4 & -2 & -1 &0 \\
\hline
$g_{\omega}$& -0.56(2)  & -0.05(2) & -1.19(2) &  1.04(2)&
1.60(4) & 2.3(2)  \\
\hline
\hline
$\tilde h_{1}$& 0.5  & 1& 1.5 & 2& 6 & 100\\
\hline
$g_{\omega}$& 0.68(12)& 0.72(15) & -0.05(5) & -0.145(3) & -0.47(2)& -0.94(2) \\
\hline
\end{tabular}
\label{tab:I}
\end{table}

\begin{table}
\caption{
Same as Table~\ref{tab:I} for case (II), i.e.,  $ h_{1}^{+}=|h_{1}|$.}
\begin{tabular}{|c|c|c|c|c|c|c|}
\hline
$\tilde h_{1}$ & -100     &-8 & -4 & -2 & -1 &0 \\
\hline
$g_{\omega}$& -0.58(2)  & -0.14(2) & -0.25(2) &  0.18(2)&
0.10(3) & 0.70(3)  \\
\hline
\hline
$\tilde h_{1}$& 0.5  & 1& 1.5 & 2& 6 & 100\\
\hline
$g_{\omega}$& 2.05(10)& 0.51(10) & 0.97(13) & 1.30(5) & -0.06(2)& -0.95(2) \\
\hline
\end{tabular}
\label{tab:II}
\end{table}

\begin{table}
\caption{
Same as Table~\ref{tab:I} for case (III), i.e.,  $h_{1}^{+}=0$.}
\begin{tabular}{|c|c|c|c|c|c|c|}
\hline
$\tilde h_{1}$ & 100     &6  & 2 & 1 &0.5 &0 \\
\hline
$g_{\omega}$& 2.8(1)  & 1.6(1) & -0.12(5) &  1.51(5)&
1.87(4) & 1.30(3)  \\
\hline
\end{tabular}
\label{tab:III}
\end{table}

%

%
\begin{figure}[h]
\includegraphics[width=0.45\textwidth]{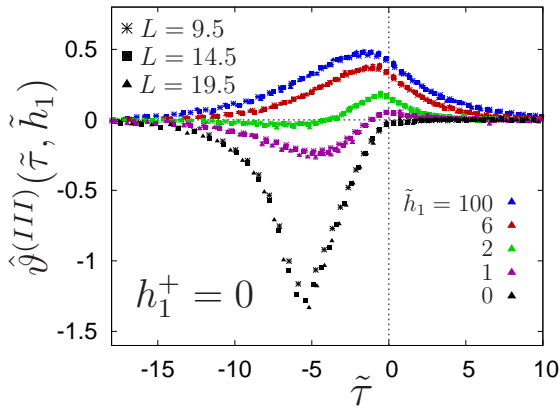}
\caption{
Same as Fig.~\ref{fig:thetat_hi} 
 for  case  (III) ($\tilde h_1\equiv \tilde h_1^-$ see Fig.~\ref{fig:fig1}). 
}
\label{fig:thetat_ho}
\end{figure}


For a selection of values of the surface field  scaling variable $\tilde h_1$
in Fig.~\ref{fig:thetat_hi}(a) we present our results for  the critical Casimir 
force scaling function $\hat \vartheta^{(I)}(\tilde \tau ,\tilde h_{1})$ corresponding to  case (I). 
We find that de facto for  $\tilde h_1=\pm 100$ the scaling
 limit of infinitely strong  surface
fields has been reached and the scaling function of the 
critical Casimir force corresponds to the fixed-point BCs $(+,\pm)$.
 In Fig.~\ref{fig:thetat_hi}(b)  
we present our  results for small values of   
$|\tilde h_1|$,  $\tilde h_{1}\in [-1,2]$,
for which  we observe a  transition from a repulsive to an attractive force
upon increasing the temperature scaling field $\tilde \tau$. In these instances in which a  change 
in sign is observed in Fig.~\ref{fig:thetat_hi}(b), the length scales
associated with the surface fields are $\ell_1^+=0$ (see Eq.~(\ref{eq:length}) for $H_1^+=\infty$) 
on the top
and  
$|\ell_1^-/\xi_{0}^+|= 0.44L_z$  (see  Eqs.~(\ref{eq:length}) and (\ref{eq:sch}) for  $ h_1=1$ and $c=0.5/\xi_0^+$)
 and $|\ell_1^-/\xi_{0}^+|\simeq 1.13L_z$ (see Eqs.~(\ref{eq:length}) and  (\ref{eq:sch}) for  $ h_1=0.5$ and $c=0.5/\xi_0^+$) 
on the bottom. (In $D=3$, 
$c=1/a $ if the coupling constant  in the surface row is unchanged relative to the one
in the bulk \cite{binder:83:0,diehl:86:0}; with $\xi_0^+\simeq 0.501$ this implies  $c\xi_{0}^+\simeq 0.501$.)

In Fig.~\ref{fig:thetat_hh}(a) we 
show data  for  the critical Casimir 
force scaling function $\hat \vartheta^{(II)}(\tilde \tau,\tilde h_{1})$ corresponding to  case (II).
For  $\tilde h_1= 100$ and $\tilde h_1= -100$  we recover the scaling limit corresponding
to $(+ +)$ and  $(+ -)$ fixed-point  BCs, respectively.
We note that the change in sign of the critical Casimir force upon varying
 the temperature occurs only for opposing surface fields. As before this change  of sign
is observed for
weak surface fields for which  $|\ell_1^-/\xi_{0}^+|\approx  L_z$, i.e.,
 for $\tilde h_1= -2, -1$.

Finally, 
 in Fig.~\ref{fig:thetat_ho} the data for  case (III) are presented. 
This case contains in particular two fixed-point  BCs:  for the value   $\tilde h_1=100$ we
observe a universal behavior of the scaling function of the  critical Casimir force corresponding
to  $(O,+)$ fixed-point BCs, whereas for  $\tilde h_1=0$
we find the  $(O,O)$ fixed-point universal behavior of  $\hat \vartheta(\tilde \tau,\tilde h_{1})$.
As in the other cases, the crossover from attraction to repulsion can be achieved by increasing
the temperature scaling variable  $\tilde \tau$, provided  that the surface fields are sufficiently weak
so that $|\ell_1^-/\xi_{0}^+|\approx L_z$, i.e., for $\tilde h_1=2$ and 1.

\section{Summary and Conclusions}
\label{sec:concl}

For variable surface fields 
we have  determined  via MC simulations the universal scaling functions 
$\hat \vartheta$ of critical Casimir forces for $3D$ Ising slabs 
describing the crossover from 
the ordinary to the normal surface universality class (Figs.~\ref{fig:thetat_hi}, \ref{fig:thetat_hh}, and
\ref{fig:thetat_ho}).
This amounts to investigate the scaling functions $\vartheta(\tau,h_1^+,h_1^-)$ (see Eq.~(\ref{eq:scf}))
for  finite values of the surface fields. 
We have computed  the  lattice scaling functions  $\hat\vartheta(\tilde \tau,\tilde h_1)$  along three different paths in the parameter space $(h_1^+,h_1^-)$  (see Fig.~\ref{fig:fig1}):
$\hat\vartheta^{(I)}(\tilde \tau,\tilde h_1)$ corresponding to  $h_1^+=\infty$,  
$\hat\vartheta^{(II)}(\tilde \tau,\tilde h_1)$ corresponding to  $h_1^+=|h_1^-|$, and $\hat\vartheta^{(II|)}(\tilde \tau,\tilde h_1)$ 
corresponding to $h_1^+=0$. Due to the fact that on the lattice the derivative in Eq.~(\ref{eq:def}) is replaced by a finite difference,
the scaling function $\hat\vartheta$ as  function of the  corrected scaling variables $(\tilde \tau,\tilde h_1)$ 
estimates   the leading behavior of
 $\vartheta$ as function of $(\tau,h_1^+,h_1^-)$; alternative definitions of the lattice derivative give rise to  distinct  corrections
for both the scaling function and the scaling variables.
We have focused on cases in which  upon variation of the  temperature a crossover from 
attractive  to repulsive critical Casimir force is observed.
Such a behavior is particularly interesting in view of potential application,
e.g., for  colloidal suspensions. 
We have found that a  change of sign of the critical Casimir force 
as a result of a minute change in temperature
occurs only in systems with  strongly asymmetrical surfaces, i.e., in cases
in which the two  surface fields differ significantly in magnitude.
For this phenomenon to occur at least one of the surface fields has to be weak enough such that
 the  length scale $\ell_1$ associated with the surface field $H_1$ (Eq.~(\ref{eq:length}))
is comparable with the width $L$ of the slab (see Figs.~\ref{fig:thetat_hi}(b) and \ref{fig:thetat_ho}
corresponding to  
fixed $h_1^+=\infty$ and $h_1^+=0$, respectively, and a variable second surface scaling field $h_1^-$).
We note that for such large values of $\ell_1$ the order parameter profiles near a
single wall differ significantly from the ones  corresponding to strong surface fields which 
belong to the surface universality class of the normal transition. 
If both surface fields are weak and have the same magnitude
 they must have  opposite signs
in order to produce a change  of  sign of the critical Casimir force (see Fig.~\ref{fig:thetat_hh}(b)).
The change
from attraction to repulsion (i.e., a zero of $\hat \vartheta$)   can occur either below the bulk critical temperature,
as for the cases in which one of the surfaces is subjected to the $(O)$  fixed-point BC or  for weak  opposing surface fields
 (see Figs.~\ref{fig:thetat_ho} and \ref{fig:thetat_hh}, respectively), or above $T_c$,
 as for the $(+)$ fixed-point BC (see Fig.~\ref{fig:thetat_hi}(b)).
In all cases the change of sign takes place rather close to the critical point.

Corrections to scaling have had to be taken into 
account in order to obtain data collapse which allowed us to infer the universal scaling functions
(see Fig.~\ref{fig:fc}). 
 The introduction  of an effective width $L+\delta$ of the slab turned out to be a very useful way of implementing
corrections to scaling, provided the surface fields are not too weak. The value of the length $\delta$ has been obtained
from the data for the critical Casimir force at the critical point (see Figs.~\ref{fig:fchi} 
and \ref{fig:fcho} corresponding to 
fixed values  $h_1^+=\infty$ and $h_1^+=0$ for the top surface, respectively, and 
a variable  surface scaling field $h_1^-$ for the bottom surface, and 
Fig.~\ref{fig:fchh} corresponding to the surface fields for the two surfaces to be of  the same magnitude; compare Fig.~\ref{fig:fig1}).

The present results close an important  gap in  
the knowledge of the Casimir scaling function for the $3D$ Ising universality class. 
The  theoretical results for  {\it  variable}
surface fields have been available in $D=2$ (from exact calculations in Ising strips
 \cite{abraham_maciolek})  and  in $D=4$ (from 
a field-theoretic approach \cite{MMD}).
The MC simulation results in  Ref.~\cite{hasen2} have been obtained for the $3D$ Blume-Capel
model which is an extension of the Ising model studied here. 
They provide 
critical Casimir forces as  function of $\beta_c-\beta$
 for  certain values of the surface  fields,   which  in the parameter space shown in Fig.~\ref{fig:fig1} correspond  to 
path $(I)$ with $h_1^-\ge 0$. 
 Because  the choice of the surface fields for the presented data is different from ours, we cannot make a direct quantitative
comparison (except for the case of ($h_1^+=\infty,h_1^-=0$) for which the data agree). However, there is 
 a qualitative agreement with our findings;
for certain choices of the  surface fields  the critical Casimir force changes sign as  function of temperature. 
This agreement provides further evidence  for the universal character  of critical Casimir forces.

Our data for the critical Casimir scaling function have the crucial  advantage over the results
in $D=2$ and 4  that they can be directly compared with possible  experimental data. 
Interestingly, in all  spatial dimensions studied 
the  crossover behavior of the  scaling function of the critical 
Casimir force as a function of  the temperature  scaling variable is qualitatively the same.
The   robustness of this observation indicates that an experimental observation of the change  of  sign
of the critical Casimir force with  temperature is possible, provided
that the chemical properties of the confining surfaces are carefuly chosen.

%

\vfill\eject

\end{document}